\documentclass[aps,reprint,showpacs,superscriptaddress,floatfix]{revtex4-1}
\pdfoutput=1
\usepackage{graphicx}
\usepackage{amsmath}
\usepackage{amssymb}
\usepackage{amsfonts}
\usepackage{color}
\usepackage{hyperref}
\usepackage{multirow}
\usepackage[usenames,dvipsnames]{xcolor}
\usepackage[caption = false]{subfig}

\newcommand{\MPQ}{Max-Planck-Institut f\"ur Quantenoptik, Hans-Kopfermann-Str.\ 1, D-85748 Garching, Germany}
\newcommand{\TU}{Department of Informatics, Technical University of Munich, Boltzmannstr.\ 3, D-85748 Garching, Germany}
\DeclareMathOperator{\Tr}{Tr}

\begin{document}

\title{Neural-Network Quantum States, String-Bond States,\\ and Chiral Topological States}

\begin{abstract}
Neural-Network Quantum States have recently been introduced as an Ansatz for describing the wave function of quantum many-body systems. We show that there are strong connections between Neural-Network Quantum States in the form of Restricted Boltzmann Machines and some classes of Tensor-Network states in arbitrary dimensions. In particular we demonstrate that short-range Restricted Boltzmann Machines are Entangled Plaquette States, while fully connected Restricted Boltzmann Machines are String-Bond States with a nonlocal geometry and low bond dimension. These results shed light on the underlying architecture of Restricted Boltzmann Machines and their efficiency at representing many-body quantum states. String-Bond States also provide a generic way of enhancing the power of Neural-Network Quantum States and a natural generalization to systems with larger local Hilbert space. We compare the advantages and drawbacks of these different classes of states and present a method to combine them together. This allows us to benefit from both the entanglement structure of Tensor Networks and the efficiency of Neural-Network Quantum States into a single Ansatz capable of targeting the wave function of strongly correlated systems. While it remains a challenge to describe states with chiral topological order using traditional Tensor Networks, we show that, because of their nonlocal geometry, Neural-Network Quantum States and their String-Bond States extension can describe a lattice Fractional Quantum Hall state exactly. In addition, we provide numerical evidence that Neural-Network Quantum States can approximate a chiral spin liquid with better accuracy than Entangled Plaquette States and local String-Bond States. Our results demonstrate the efficiency of neural networks to describe complex quantum wave functions and pave the way towards the use of String-Bond States as a tool in more traditional machine-learning applications.
\end{abstract}


\author{Ivan Glasser}
\affiliation{\MPQ}
\author{Nicola Pancotti}
\affiliation{\MPQ}
\author{Moritz August}
\affiliation{\TU}
\author{Ivan D. Rodriguez}
\affiliation{\MPQ}
\author{J. Ignacio Cirac}
\affiliation{\MPQ}

\maketitle

\section{Introduction}
\label{SEC:Introduction}

\definecolor{darkblue}{rgb}{0.0, 0.0, 0.55}

Recognizing complex patterns is a central problem that pervades all fields of science. The increased computational power of modern computers has allowed the application of advanced methods to the extraction of such patterns from humongous amounts of data and we are witnessing an ever increasing effort to find novel applications in numerous disciplines. This led to a line of research now called Quantum Machine Learning\cite{BiamonteReview}, which is divided in two main different branches. The first tries to develop quantum algorithms capable of learning, i.e. to exploit speed ups from quantum computers to make machines learn faster and better. The second, that we will consider in this work, tries to use classical machine learning algorithms to extract insightful information about quantum systems. 

The versatility of machine learning has allowed scientists to employ it in a number of problems which span from quantum control\cite{ZahedinejadQuantumControl,MoritzPaper, BanchiQuantumControl} and error correcting codes\cite{TorlaiDecoder} to tomography \cite{Torlai2017}. In the last few years we are experiencing interesting developments also for some central problems in condensed matter, such as quantum phase classification/recognition\cite{phasesByConfusion, Carrasquilla2017, Broecker2017,Wang2016}, improvement of dynamical mean field theory\cite{Arsenault2014}, enhancement of Quantum Monte Carlo methods \cite{self-learningMonteCarlo, acceleratedMonteCarlo} or approximations of thermodynamic observables in statistical systems\cite{ThermodynamicsRBM}.

An idea which received a lot of attention from the scientific community consists in using neural networks as variational wave functions to approximate ground states of many-body quantum systems\cite{Carleo2016}. These networks are trained/optimized by the standard Variational Monte Carlo (VMC) method and while a few different neural-network architectures have been tested\cite{Carleo2016,Cai2017,Saito2017}, the most promising results so far have been achieved with Boltzmann Machines\cite{Hinton1985}. In particular, state of the art numerical results have been obtained on popular models with Restricted Boltzmann Machines (RBM) and recent effort has demonstrated the power of Deep Boltzmann Machines to represent ground states of many-body Hamiltonians with polynomial-size gap and quantum states generated by any polynomial size quantum circuits\cite{Gao2017,Huang2017}.

Other seemingly unrelated classes of states that are widely used in condensed matter physics are Tensor Networks States. In 1D, Matrix Product States (MPS) can approximate ground states of physical Hamiltonians efficiently \cite{Hastings2007,Verstraete2006} and their structure has led to both analytical insights over the entanglement properties of physical systems as well as efficient variational algorithms for approximating them\cite{VerstraeteReview,SchollwockReview, DMRG_white}. The natural extension of MPS to larger dimensional systems are Projected Entangled Pair States (PEPS)\cite{peps2004}, but their exact contraction is $\#P$ hard\cite{Schuch2007} and algorithms for optimizing them need to rely on approximations. Another approach to define higher dimensional Tensor Networks consists in first dividing the lattice into overlapping clusters of spins. The wave function of the spins in each cluster is then described by a simple Tensor Network. The global wave function is finally taken to be the product of these Tensor Networks, which introduces correlations among the different clusters. This construction for local clusters parametrized by a full tensor gives rise to Entangled Plaquette States (EPS)\cite{Gendiar2002,Mezzacapo2009,Changlani2009}, while taking one dimensional clusters of spins each described by a MPS leads to a String-Bond States (SBS) Ansatz\cite{SBS2008,Sfondrini2010}. These states can be variationally optimized using the VMC method\cite{Sandvik2007,SBS2008} and have been applied to 2D and 3D systems.

All these variational wave functions have been successful in describing strongly correlated quantum many body systems, including topologically ordered states. The Toric code\cite{Kitaev2003} is a prototypical example which can be written exactly as a PEPS\cite{Verstraete2006b}, an EPS\cite{Changlani2009}, a SBS\cite{SBS2008} or a short-range RBM\cite{Deng2016}.  This shows that in some cases Tensor Networks and Neural-Network Quantum States can be related. Indeed it was recently shown that local Tensor Networks can be represented efficiently by Deep Boltzmann Machines\cite{Gao2017,Huang2017,Chen2017}. Not every topological state can however easily be represented by local Tensor Networks. A class of states for which this is challenging are chiral topological states breaking time-reversal symmetry. Such states were first realized in the context of the Fractional Quantum Hall (FQH) effect\cite{Tsui1982} and significant progress has since been made towards the construction of lattice models displaying the same physics, either in Hamiltonians realizing fractional Chern insulators\cite{Levin2009,Sheng2011,Neupert2011,Wang2011,Sun2011,Regnault2011} or in quantum anti-ferromagnets on several lattices\cite{Nielsen2013,Bauer2014,He2014,Gong2014}. One approach to describe the wave function of these anti-ferromagnets is to use parton constructed wave functions\cite{Baskaran1987,Affleck1988,Wen1999,Hu2015}. It has also been suggested to construct chiral lattice wave functions from the FQH continuum wave functions, the paradigmatic example being the Kalmeyer-Laughlin wave function\cite{Kalmeyer1987}. Efforts to construct chiral topological states with PEPS have been undertaken recently\cite{Thorsten2013,Dubail2015,Shuo2015,Poilblanc2015, YANG_chiral_PEPS}, but the resulting states are critical. In the non-interacting case it has moreover been proven  that the local parent Hamiltonian of a chiral fermionic Gaussian PEPS has to be gapless\cite{Dubail2015}.

In this work we show that there is a strong relation between Restricted Boltzmann Machines and Tensor Network States in arbitrary dimension. We demonstrate that short-range RBM are a special subclass of EPS, while fully-connected RBM are a subclass of SBS with a flexible non-local geometry and low bond dimension. This relation provides additional insights over the geometric structure of RBM and their efficiency. We discuss the advantages and drawbacks of RBM and SBS and provide a way to combine them together. This generalization in the form of non-local String-Bond States takes leverage of both the entanglement structure of Tensor Networks and the efficiency of RBM. It allows for the description of states with larger local Hilbert space and has a flexible geometry. It can moreover be combined with more traditional Ansatz wave functions that serve as an initial approximation of the ground state.

We then apply these methods to the challenging problem of approximating chiral topological states. We prove that any Jastrow wave function, and thus the Kalmeyer-Laughlin wave function, can be written exactly as a RBM. We moreover show that a remarkable accuracy can be achieved numerically with much less parameters than is required for an exact construction. We numerically evaluate the power of EPS, SBS and RBM to approximate the ground state of a chiral spin liquid for which the Laughlin state is already a good approximation\cite{Nielsen2013} and find that RBM and non-local SBS are able to achieve lower energy than the Laughlin wave function. By combining these classes of states with the Laughlin wave function, we are able to reach even lower energies and to characterize the properties of the ground state of the model.

The paper is organized as follows: in Section~\ref{section1} we introduce the Variational Monte Carlo method and how it can be used to optimize both Tensor-Network and Neural-Network States. In Section~\ref{section2} the mapping between RBM, EPS and SBS is derived and its geometric implications are discussed. Finally we apply these techniques to the approximation of chiral topological states in Section~\ref{section3}.

\section{Variational Monte Carlo with Tensor Networks and Neural-Network States}
\label{section1}

\subsection{The Variational Monte Carlo method}

Given a general Hamiltonian $H$, one of the main challenges of quantum many-body physics is to find its ground state $|\psi_0\rangle$ satisfying the Schr\"{o}dinger equation $H|\psi_0\rangle=E_0|\psi_0\rangle$. This eigenvalue problem can be mapped to an optimization problem through the variational principle, stating that the energy of any quantum state is higher than the energy of the ground state. A general pure quantum state on a lattice with $N$ spins can be expressed in the basis spanned by $|s_1,\ldots,s_N\rangle$, where $s_i$ are the projection of the spins on the z axis, as 
\begin{align}
|\psi\rangle=\sum_{s_1,\ldots,s_N}\psi(s_1,\ldots,s_N)|s_1,\ldots,s_N\rangle.
\end{align}
Finding the ground state amounts to finding the exponentially many parameters $\psi(s_1,\ldots,s_N)$ minimizing the energy, which can only be done exactly for small sizes. Instead of searching for the ground state in the full Hilbert space, one may restrict the search to an Ansatz class specified by a particular form for the function $\psi_w(s_1,\ldots,s_N)$ depending on polynomially many variational parameters $w$. The Variational Monte Carlo method \cite{McMillan1965,ReviewVMC} (VMC) provides a general algorithm for optimizing the energy of such a wave function. One can compute the energy by expressing it as 
\begin{align}
E_w=\frac{\langle\psi |H|\psi\rangle}{\langle\psi|\psi\rangle}=\sum_\mathbf{s}  p(\mathbf{s}) E_{\text{loc}}(\mathbf{s}),
\end{align}
where $\mathbf{s}=s_1,\ldots,s_N$ is a spin configuration, $p(\mathbf{s})=\frac{|\psi_w(\mathbf{s})|^2}{\sum_\mathbf{s} |\psi_w(\mathbf{s})|^2}$ is a classical probability distribution and the local energy $E_{\text{loc}}(\mathbf{s})=\sum_{\mathbf{s}'}\langle \mathbf{s}|H|\mathbf{s}'\rangle \frac{\psi_w(\mathbf{s}')}{\psi_w(\mathbf{s})}$ can be evaluated efficiently for Hamiltonians involving few-body interactions. The energy is therefore an expectation value with respect to a probability distribution $p$ that can be evaluated using Markov Chain Monte Carlo sampling techniques such as the Metropolis-Hastings algorithm \cite{Metropolis1953,Hastings1970}. The second ingredient required to minimize the energy with respect to the parameters $w$ is the gradient of the energy, which can be expressed in a similar form since
\begin{align}
\frac{\partial E_w}{\partial w_i}=2 \sum_\mathbf{s}  p(\mathbf{s}) \Delta_{w_i}(\mathbf{s})^* \left(E_{\text{loc}}(\mathbf{s})-E_w\right),
\end{align}
where we have defined $\Delta_{w_i}(\mathbf{s})=\frac{1}{\psi_w(\mathbf{s})}\frac{\partial \psi_w(\mathbf{s})}{\partial w_i}$ as the log-derivative of the wave function with respect to some parameter $w_i$. This is also an expectation value with respect to the same probability distribution $p$ and can therefore be sampled at the same time, which allows for the use of gradient-based optimization methods. At each iteration, the energy and its gradient are computed with Monte Carlo, the parameters $w$ are updated by small steps in the direction of negative energy derivative ($w_i \leftarrow w_i-\alpha \frac{\partial E_w}{\partial w_i}$) and the process is repeated until convergence of the energy. The VMC method, in its simplest form, only requires the efficient computation of $\frac{\psi_w(\mathbf{s}')}{\psi_w(\mathbf{s})}$ for two spin configurations $s$ and $s'$, as well as the log-derivative of the wave function $\Delta_w(\mathbf{s})$. More efficient optimization methods can be used, such as conjugate-gradient descent, Stochastic Reconfiguration\cite{Sorella2001,Sorella2005}, the Newton method\cite{Umrigar2005} or the linear method\cite{Nightingale2001,Toulouse2007,Umrigar2007}.

At this point one has to choose a special form for the wave function $\psi_w$. One of the traditional variational wave functions for a many-body quantum system is a Jastrow wave function\cite{McMillan1965,Jastrow}, which consists in its most general form of a product of wave functions for all pairs of spins:
\begin{align}
\psi_w(\mathbf{s})=\prod_{i<j}f_{ij}(s_i,s_j),\label{jastrowavefunction}
\end{align}
where each $f_{ij}$ is fully specified by its four values $f_{ij}(s_i,s_j),\ s_i,s_j\in\{-1,1\}$. Such an Ansatz does not presuppose a particular local geometry of the many-body quantum state: in general this Ansatz can be non-local due to the correlations between all pairs of spins (Fig.~\ref{jastrow}). A local structure can be introduced by choosing a form for $f_{ij}$ which decays with the distance between position $i$ and $j$.

\subsection{Variational Monte Carlo method with Tensor Networks}
\label{vmc}

In condensed matter physics, important assets to simplify the problem are the geometric structure and locality of physical Hamiltonians. In 1D, it has been proven that ground states of gapped local Hamiltonians have an entanglement entropy of a subsystem which grows only like the boundary of the subsystem\cite{Hastings2007}. States satisfying such an area-law can be efficiently approximated by Matrix Product States (MPS)\cite{Verstraete2006}. Matrix Product State are one dimensional Tensor Network States whose wave function for a spin configuration reads
\begin{align}
\psi_w(\mathbf{s})=\Tr \left(\prod_{j=1}^N A^{s_j}_j\right).
\end{align}
For each spin and lattice site, the matrix $A_i^{s_i}$ of dimension $D\times D$, where $D$ is called the bond dimension, contains the variational parameters. Matrix Product States can be efficiently optimized using the Density Matrix Renormalization Group (DMRG)\cite{White1992}, but the previously described VMC method can also be applied\cite{Sandvik2007,SBS2008} by observing that the ratio of two configurations is straightforward to compute, and that the log-derivative with respect to some matrix $A_k^{s_k'}$ is given by
\begin{align}
\Delta_{A_k^{s_{k'}}}(\mathbf{s})=\frac{\delta_{s_k,s_{k'}} \left(A^{s_k}_{k+1}\cdots A^{s_N}_N A^{s_1}_1 A^{s_{k-1}}_{k-1}\right)^\top}{\Tr(A^{s_1}_1\cdots A^{s_N}_N)}.\label{logdMPS}
\end{align}
In some cases, this method is less likely to be trapped in a local minimum than DMRG, since all coefficients can be updated at once. In addition, the cost only scales as $O(D^3)$ in the bond dimension for periodic boundary conditions.

\begin{figure}[t]
\centering
\subfloat[Jastrow]{\includegraphics[width = 3cm]{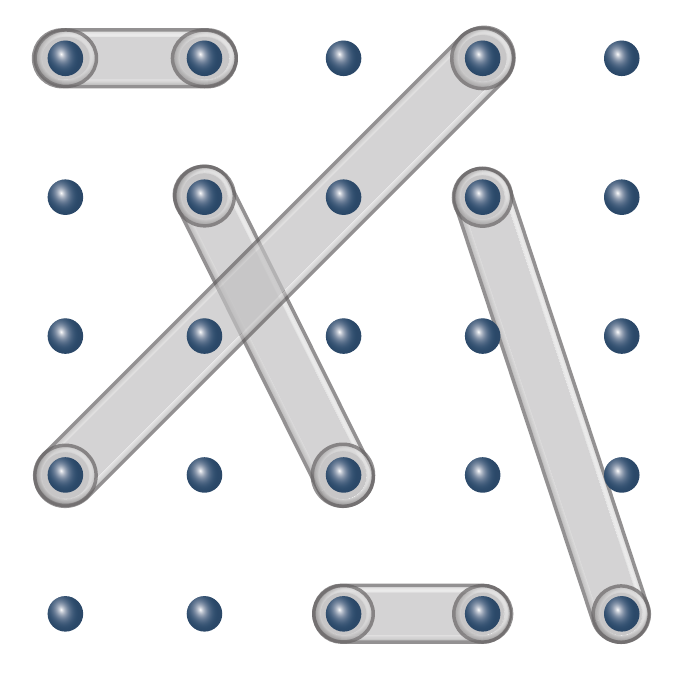}
\label{jastrow}} 
\subfloat[MPS]{\includegraphics[width = 3cm]{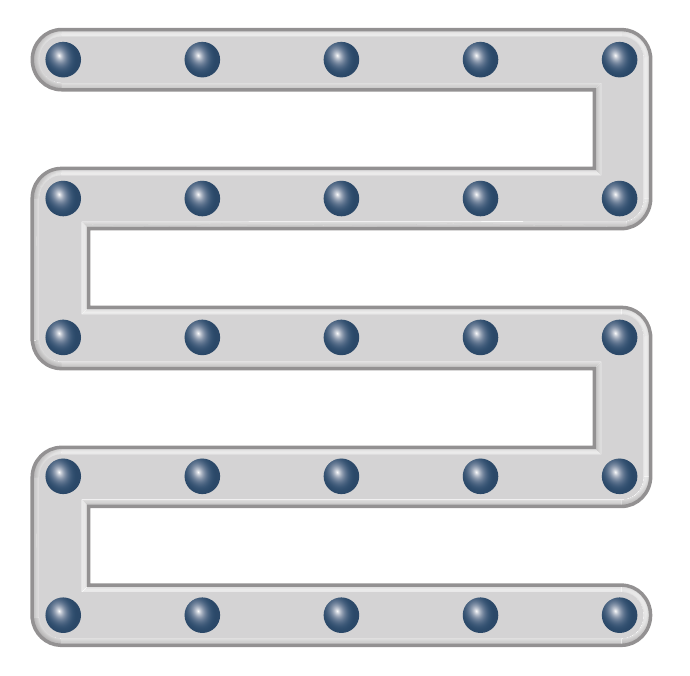} \label{mps}} \\
\subfloat[EPS]{\includegraphics[width = 3cm]{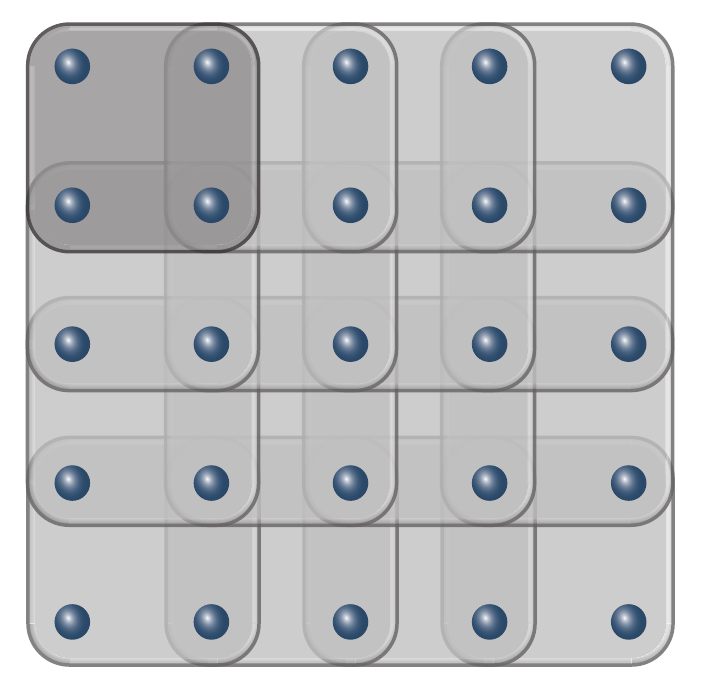} \label{eps}}
\subfloat[SBS]{\includegraphics[width = 3cm]{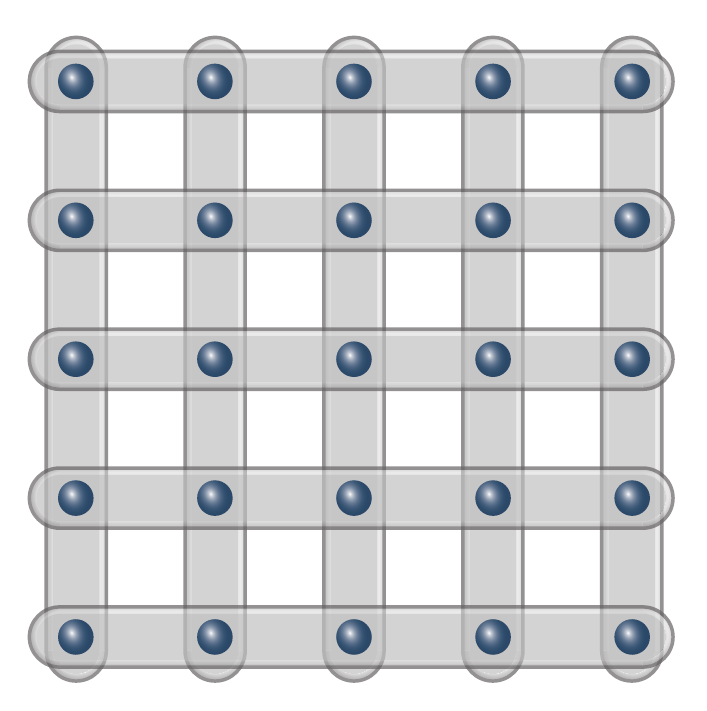} \label{sbs}}
\caption{Geometry of Ansatz wave functions: (a) Jastrow wave function include correlations within all pairs of spins. (b) Matrix Product States (MPS) in 2D cover the lattice with one snake. (c) Entangled Plaquette States (EPS) include all spin correlations within each plaquette (2x2 on the figure) and mediate correlations between distant spins through overlapping plaquettes.  (d) String-Bond States (SBS) cover the lattice with many 1D strings on which the interactions within spins are captured by a MPS.}
\end{figure}

In higher dimensions, Matrix Product States can be defined by mapping the system to a line (Fig.~\ref{mps}). The problem of this construction is evident from Fig.~\ref{mps}. Spins which sit close to each other might be separated by a long distance on the line, the Ansatz thus fails to reproduce the local structure of the state, which leads to an exponential scaling of the computing resources needed with the system size\cite{Liang1994}. The natural extension of MPS to 2D systems are Projected Entangled Pair States (PEPS)\cite{peps2004}, for which the wave function can be written as a contraction of local tensors on the 2D lattice. While PEPS have been successful in describing strongly correlated quantum many body systems, their exact contraction is $\#P$ hard\cite{Schuch2007} and their optimization cannot rely on the standard VMC method without approximations. In the following we will instead consider other classes of tensor-network states in more than one dimension for which the exact computation of the wave function is efficient, which allows for the direct use of the VMC method.

One approach consists in cutting a lattice in $P$ small clusters of $n_p$ spins, or plaquettes, and construct the wave function exactly on each plaquette. The wave function of the full quantum system is then taken to be the product of the wave functions in each plaquette, in a mean-field fashion. Choosing overlapping plaquettes allows one to go beyond mean-field and include correlations between different plaquettes (Fig.~\ref{eps}). The wave function of such an Entangled Plaquette State (EPS, also called a Correlated Product State) is written as\cite{Gendiar2002,Mezzacapo2009,Changlani2009}:
\begin{align}
\psi_w(\mathbf{s})=\prod_{p=1}^{P} C_p^{\mathbf{s}_p},\label{epswavefunction}
\end{align} 
where a coefficient $C_p^{\mathbf{s}_p}$ is assigned to each of the $2^{n_p}$ (for spin-$1/2$ particles) configurations $\mathbf{s}_p=s_{a_1},\ldots ,s_{a_{n_p}}$ of the spins on the plaquette $p$. Each $C_p$ can be seen as the most general function on the Hilbert space corresponding to the spins in plaquette $p$. The accuracy can be improved by enlarging the size of the plaquettes and the Ansatz is exact once the size of the plaquettes reaches the size of the lattice (which can only be achieved on small lattices). Moreover, once the spin configuration $\mathbf{s}_p$ is fixed, the log-derivative of the wave function with respect to the variational parameters is simply
\begin{align}
\Delta_{C_p^{\mathbf{s}_p}}(\mathbf{s})=\frac{1}{C_p^{\mathbf{s}_p}},\label{logdeps}
\end{align} 
which is efficient to compute.

EPS are limited to small plaquettes since for each plaquette the number of coefficients scales exponentially with the size of the plaquette. However one can generalize this Ansatz by describing the state of clusters of spins by a MPS, avoiding the exponentially many coefficients needed. The lattice is now cut in overlapping 1D strings which can mediate correlations on longer distances compared to local plaquettes (Fig.~\ref{sbs}). The resulting Ansatz is a String-Bond State (SBS)\cite{SBS2008} defined by a set of strings $i\in S$ (each string $i$ is an ordered subset of the set of spins) and a MPS for each string:
\begin{align}
\psi_w(\mathbf{s})=\prod_{i} \Tr\left(\prod_{j\in i} A^{s_j}_{i,j} \right).
\end{align} 
The descriptive power of this Ansatz is highly dependant on the choice of strings: for example, by using small strings covering small plaquettes and a large bond dimension it includes EPS; whereas a single long string in a snake pattern includes MPS in 2D. In 3D, it has been used by choosing strings parallel to the axes of the lattice\cite{Sfondrini2010}. Since the form of the wave function is a product of MPS, the log-derivative with respect to some elements present in one of the MPS is simply the log-derivative for the corresponding MPS (Eq.~\eqref{logdMPS}). The VMC procedure for optimizing SBS and MPS thus have the same cost. In addition, the ratio of two configurations which differ only by a few spins can be computed by considering only the strings including these spins, which speeds up the computation considerably. Let us note that a SBS can be mapped analytically to a MPS, but that the resulting MPS would have a bond dimension exponential in the number of strings.

\subsection{Variational Monte Carlo method with Neural Networks}

\begin{figure}[h]
\centering
\subfloat[Deep Boltzmann Machine]{\includegraphics[width = 2.5cm]{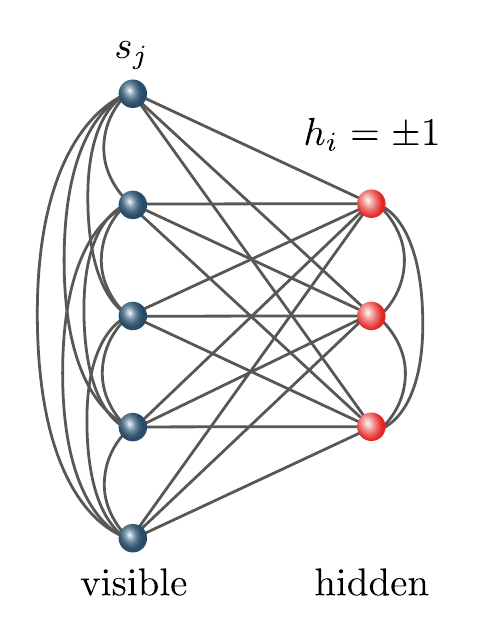} \label{dbm}}
\subfloat[Restricted Boltzmann Machine in 2D]{\includegraphics[width = 4cm]{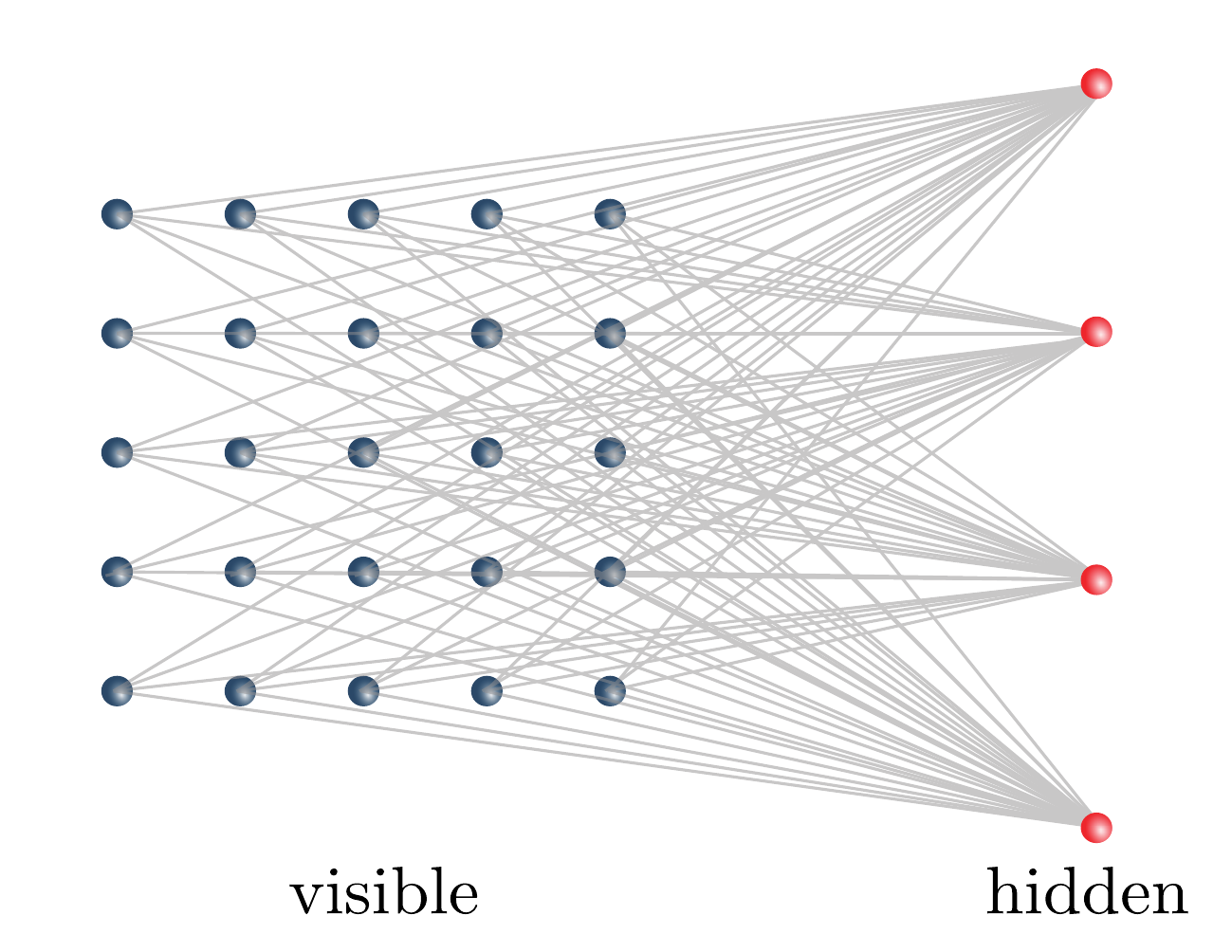} \label{rbm}}
\caption{(a) Boltzmann Machines approximate a probability distribution by the Boltzmann weights of an Ising Hamiltonian on a graph including visible units (corresponding to the spins $s_j$) and hidden units $h_i$ which are summed over. (b) Restricted Boltzmann Machines (here in 2D) only include interactions between the visible and the hidden units.}
\end{figure}

Recently, it was realized that the VMC method can be viewed as a form of learning, which motivated the use of another class of seemingly unrelated states for describing the ground state of many-body quantum states: Neural-Network Quantum States\cite{Carleo2016} are quantum states for which the wave function has the structure of an artificial neural network. While a few different networks have been investigated\cite{Carleo2016,Cai2017,Saito2017,Torlai2017}, the most promising results so far have been obtained with Boltzmann Machines\cite{Hinton1985}. Boltzmann Machines are a kind of generative stochastic artificial neural networks that can learn a distribution over the set of their inputs. In quantum many-body physics, the inputs are spin configurations and the wave function is interpreted as a (complex) probability distribution that the networks tries to approximate. Boltzmann Machines consist of two sets of binary units (classical spins): the visible units $v_i, \ i \in \{1,\ldots,N\}$, corresponding to the configurations of the original spins in a chosen basis, and hidden units $h_j, \ j \in \{1,\ldots,M\}$ which introduce correlations between the visible units. The whole system interacts through an Ising interaction which defines a joint probability distribution over the visible and hidden units as the Boltzmann weight of this Hamiltonian:
\begin{align}
P(\mathbf{v},\mathbf{h})=\frac{1}{Z}e^{\mathcal{H}(\mathbf{v},\mathbf{h})},
\end{align}
where the Hamiltonian $\mathcal{H}$ is defined as
\begin{align}
\mathcal{H}=&\sum_{j}a_j v_j + \sum_i b_i h_i+\sum_{i<j} c_{ij} v_i v_j \nonumber
\\
&+\sum_{i,j} w_{ij} h_i v_j + \sum_{i<j} d_{ij} h_i h_j,\nonumber
\end{align}
and $Z$ is the partition function. The marginal probability of a visible configuration is then given by summing over all possible hidden configurations:
\begin{align}
P(\mathbf{v})=\sum_{\mathbf{h}}\frac{1}{Z}e^{\mathcal{H}(\mathbf{v},\mathbf{h})},
\end{align}
and we take this quantity as Ansatz for the wave function: $\psi_w(\mathbf{s})=P(\mathbf{s})$. The variational parameters are the complex parameters of the Ising Hamiltonian. In the case where there are interactions between the hidden units (Fig.~\ref{dbm}), the Boltzmann Machine is called a Deep Boltzmann Machine. 
It has been shown that Deep Boltzmann Machines can efficiently represent ground states of many-body Hamiltonians with polynomial-size gap, local tensor-network state and quantum states generated by any polynomial size quantum circuits\cite{Gao2017,Huang2017,Chen2017}. On the other hand, computing the wave function $\psi_w(\mathbf{s})$ of such a Deep Boltzmann Machine in the general case is intractable, due to the exponential sum over the hidden variables, so the VMC method cannot be applied to Deep Boltzmann Machines without approximations.
We therefore turn to the investigation of Restricted Boltzmann Machines (RBM), which only include interactions between the visible and hidden units (as well as the one-body interaction terms which correspond to biases). In this case, the sum over the hidden units can be performed analytically and the resulting wave function can be written as (here we take the hidden units to have values $\pm 1$):
\begin{align}
\psi_w(\mathbf{s})=e^{\sum_{j}a_j s_j}\prod_i \cosh\left(b_i +\sum_{j} w_{ij}s_j\right).
\end{align}
RBM can represent many quantum states of interest, such as the toric code\cite{Deng2016}, any graph state, cluster states and coherent thermal states\cite{Gao2017}; the possibility of computing efficiently $\psi_w(\mathbf{s})$ prevents it however to approximate all PEPS and ground states of local Hamiltonians\cite{Gao2017}. On the other hand, since computing $\psi_w(\mathbf{s})$ and its derivative is very efficient, RBM can be optimized numerically via the VMC method.

\section{Relationship between Tensor-Network and Neural-Network states}
\label{section2}

While the machine learning perspective which leads to the application of Boltzmann Machines to quantum many-body systems seems quite different from the information-theoretic approach to the structure of tensor-network states, we will see that they are in fact intimately related. It was recently shown that while fully connected RBM can exhibit volume-law entanglement, contrary to local tensor networks, all short-range RBM satisfy an area law\cite{Deng2017}. Moreover short-range and sufficiently sparse RBM can be written as a MPS\cite{Chen2017}, but doing so for a fully-connected RBM would require an exponential scaling of the bond dimension with the size of the system. In this section we show that there is a tighter connection between RBM and the previously introduced tensor networks in arbitrary dimension.

\subsection{Jastrow wave functions, RBM and the Majumdar-Gosh model}
\label{jastrowrbmcorrespondence}

Before turning to tensor networks, let us first consider the simple case of the Jastrow wave function (Eq.~\eqref{jastrowavefunction}). Boltzmann Machines including only interactions between the visible units lead to a wave function
\begin{align}
\psi_w(\mathbf{s})=\prod_{k}e^{a_k s_k}\prod_{i<j}e^{c_{ij} s_i s_j},
\end{align} 
which has the form of a product between functions of pairs of spins, and is thus a Jastrow wave function. More generally, semi-restricted Boltzmann Machines including interactions between visible units as well as between hidden and visible units are a product of a RBM and a Jastrow factor.

Nevertheless, one may ask whether a RBM alone is enough to describe a Jastrow factor. We first rewrite the RBM as 
\begin{align}
\psi_w(\mathbf{s})=\prod_j A_j^{s_j} \prod_i \left(B_i \prod_j W_{ij}^{s_j}+\frac{1}{B_i \prod_j W_{ij}^{s_j}}\right) ,
\label{RBMlog}
\end{align}
where we have redefined the parameters with uppercase letters as the exponential of the original parameters, thus removing the exponentials in the hyperbolic cosine. This form will be convenient for the numerical simulations presented later. Since Jastrow wave functions are a product of functions of all pairs of spins, let us show that a RBM with one hidden unit can represent any function of two spins. It then follows that a RBM with $M=N(N-1)/2$ hidden units, each representing a function of one pair of spins, can represent a Jastrow wave function with polynomial resources. We thus have to solve for a system of four non-linear equations with $s_1,s_2\in\{-1,1\}$ and $f$ the most general function of two spins : $\psi_w(s_1,s_2)=f(s_1,s_2)$. This system is solved in Appendix \ref{Appjastrowasrbm}, providing an analytical solution for the parameters of the RBM to represent the Jastrow wave function exactly, or to arbitrary precision if $f(s_1,s_2)=0$ for some spins.

As an application, we use this result to write the ground state of the Majumdar-Gosh model\cite{MajumdarGosh1969} exactly as a RBM. The Majumdar-Ghosh model is defined by the following spin-$1/2$ Hamiltonian:
\begin{align}
H=J\sum_{i=1}^{N-1} \mathbf{S}_i\cdot \mathbf{S}_{i+1}+\frac{J}{2}\sum_{i=1}^{N-2} \mathbf{S}_i\cdot \mathbf{S}_{i+2}
\end{align}
The ground state wave function is a product of singlets formed by neighboring pairs of spins:
\begin{align}
|\psi\rangle\propto\prod_{n=1}^{N/2}|\uparrow_{2n-1}\rangle|\downarrow_{2n}\rangle-|\downarrow_{2n-1}\rangle|\uparrow_{2n}\rangle,
\end{align}
This wave function can also be expanded in the computational basis as
\begin{align}
\psi(s_1,\ldots,s_N)&\propto\prod_{n=1}^{N/2}(-1)^{(s_{2n-1}+3)/2}\delta_{s_{2n-1}\neq s_{2n}},\\
&\propto \prod_{n=1}^{N/2} f(s_{2n-1},s_{2n}).
\end{align}
Using the previous result, each function of two spins $f$ can be written as a RBM using one hidden unit, which leads to a RBM representation of the ground states with $M=N/2$ hidden units. We also find numerically on small systems that a RBM using less than $M=N/2$ has higher energy than the ground state, which suggests that $M=N/2$ could be optimal.

\subsection{Short-range RBM are EPS}

\begin{figure}[t]
\centering
\subfloat[Local RBM as an EPS]{\includegraphics[width = 3.5cm]{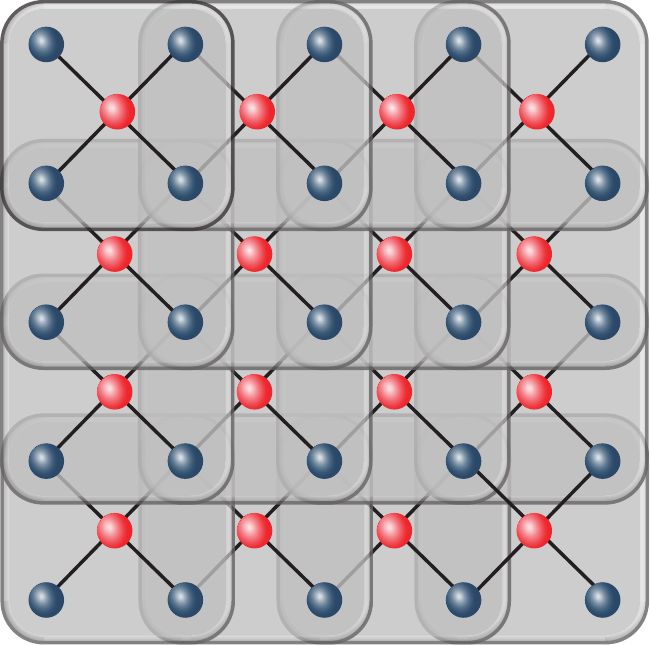}\label{rbmeps}}
\subfloat[RBM as a non-local SBS]{\includegraphics[width = 4cm]{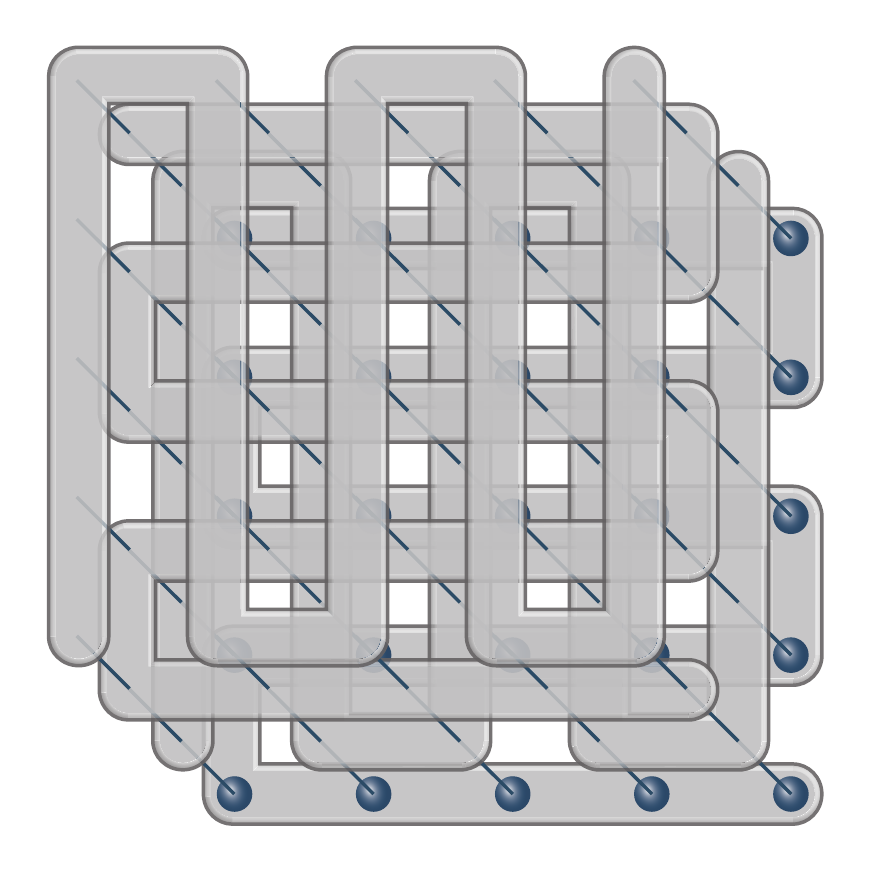}\label{rbmsbs}}
\caption{(a) A locally connected RBM is an EPS where each plaquette encodes the local connections to a hidden unit. (b) Once expressed as a SBS a fully-connected RBM can be represented by many strings on top of each other. Enlarging the RBM by using non-commuting matrices to non-local SBS induces a geometry in each string.}
\end{figure}

Let us now turn to the specific case of RBM with short-range connections (sRBM). This encompasses all quantum states that have previously been written exactly as a RBM, such as for example the toric code or the 1D symmetry-protected topological cluster state\cite{Deng2016}. Such states have weights connections between visible hidden units that are local. Each hidden unit is connected to a local region with at most $d$ neighboring spins. If we divide the lattice into $M$ subsets $p_i,\ i\in\{1,\ldots,M\}$, the wave function can be rewritten as (we omit here the biases $a_j$ which are local one-body terms):
\begin{align}
\psi_w(\mathbf{s})&=\prod_{i=1}^M \cosh\left(b_i +\sum_{j\in p_i} w_{ij}s_j\right)\\
&=\prod_{i=1}^M C_i^{\mathbf{s}_i},
\end{align}
where $\mathbf{s}_i$ is the spin configuration in the subset $p_i$. This is the form (Eq.\eqref{epswavefunction}) of an EPS (Fig.~\ref{rbmeps}). For translational invariant systems, the short-range RBM becomes a convolutional RBM, which corresponds to a translational invariant EPS. The main difference between a short-range RBM and an EPS is that the RBM considers a very specific function among all possible functions of the spins inside a plaquette, hence EPS are more general than short-range RBM. This also directly implies that the entanglement of short-range RBM follows an area law. The main advantage of short-range RBM over EPS is that due to the exponential scaling of EPS with the size of the plaquettes, larger plaquettes can be used in short-range RBM than in EPS. Since in practice for finite systems it is possible to work directly with fully-connected RBM, we argue that EPS or fully-connected RBM should be preferred to short-range RBM for numerical purposes.

\subsection{Fully-connected RBM are SBS}

Fully-connected RBM, on the other hand, do not always satisfy an area law\cite{Deng2017} and hence cannot always be approximated by local tensor networks. Nevertheless, one can express the RBM wave function as (here we also omit the bias $a_j$):
\begin{align}
\psi_w(\mathbf{s})&=\prod_i \cosh\left(b_i +\sum_{j} w_{ij}s_j\right)\\
&\propto \prod_i \left(e^{b_i+\sum_j w_{ij} s_j}+e^{-b_i-\sum_j w_{ij} s_j}\right)\\
&\propto \prod_i \Tr \begin{pmatrix}
e^{b_i+\sum_j w_{ij} s_j} & 0 \\
0 & e^{-b_i-\sum_j w_{ij} s_j}
\end{pmatrix}\\
&\propto \prod_i \Tr \left(\prod_{j\in i} A^{s_j}_{i,j}\right),
\end{align}
where
\begin{align}
A^{s_j}_{i,j}=\begin{pmatrix}
e^{b_i/N+w_{ij} s_j} & 0 \\
0 & e^{-b_i/N-w_{ij} s_j}
\end{pmatrix}
\end{align}
are diagonal matrices of bond dimension $2$. This shows that RBM are String-Bond States, as the wave function can be written as a product of MPS over strings, where each hidden unit corresponds to one string. The only difference between the SBS as depicted in Fig.~\ref{sbs} and the RBM is the geometry of the strings. In a fully-connected RBM, each string goes over the full lattice, while SBS have traditionally been used with smaller strings and with at most a few strings overlapping at each lattice site.

\subsection{Generalizing RBM to non-local SBS}

In the SBS language, RBM consists in many strings overlapping on the full lattice. The matrices in each string in the RBM are diagonal, hence commute, so they can be moved in the string up to a reordering of the spins. This means that each string does not have a fixed geometry and can adapt to stronger correlations in different parts of the lattice, even over long distances. This motivates us to generalize RBM to SBS with diagonal matrices in which each string covers the full lattice (Fig.~\ref{rbmsbs}). In the following we denote these states as non-local dSBS. This amounts to relaxing the constraints on the RBM parameters to the most general diagonal matrix and enlarging the bond dimension of the matrices. For example taking the matrices
\begin{align}
A^{s_j}_{i,j}=\begin{pmatrix}
a^{s_j}_{i,j} & 0 & 0 \\
0 & b^{s_j}_{i,j}& 0\\
0 & 0 & c^{s_j}_{i,j}
\end{pmatrix},
\end{align}
with different parameters $a^{s_j}_{i,j}$ for each string, lattice site and spin direction, leads to the wave function (here $D=3$):
\begin{align}
\psi_w(\mathbf{s})&=\prod_i \left( \prod_j a^{s_j}_{i,j} + \prod_j b^{s_j}_{i,j} + \prod_j c^{s_j}_{i,j} \right).\label{spin1rbm}
\end{align}
Note that even for $2\times 2$ matrices, the non-local dSBS is more general than a RBM since the coefficients in each of the two matrices corresponding to one spin are independent from each other, which is not the case in the RBM.

Generalizing such a wave function to larger spins than spin-$1/2$ is straightforward, since the spin $s_i$ is just indexing the parameters. This provides a way of defining a natural generalization of RBM which can handle systems with larger physical dimension. For instance this can be applied to spin-1 systems, while a naive construction for a RBM with spin-1 visible and hidden units leads to additional constraints, as well as to approximate bosonic systems by truncating the local Hilbert space of the bosons.

A further way to extend this class of states is to include non-commuting matrices. This fixes the geometry of each string by defining an order and also enables to represent more complicated interactions. In the following we will refer to SBS in such a geometry as non-local SBS. The advantage of this approach is that it can capture more complex correlations within each string, while introducing additional geometric information about the problem at hand. It comes however at a greater numerical cost than  non-local dSBS or RBM due to the additional number of parameters. In practice, one can use an already optimized RBM or dSBS as a way of initializing a non-local SBS.

In some cases, the SBS representation is more compact than the RBM/dSBS representation. Let us consider again the ground state of the Majumdar-Gosh Hamiltonian, which we previously wrote as a RBM with $M=N/2$ hidden units. The ground state of the Majumdar-Gosh Hamiltonian can also be written as a simple MPS with bond dimension 3 and periodic boundary conditions, with matrices \cite{SchollwockReview}
\begin{align}
A_n^{s_n=-1}=\begin{pmatrix}
0 & 1 & 0 \\
0 & 0 & -\frac{1}{\sqrt{2}}\\
0 & 0 & 0
\end{pmatrix}, 
A_n^{s_n=1}=\begin{pmatrix}
0 & 0 & 0 \\
\frac{1}{\sqrt{2}} & 0 & 0\\
0 & 1 & 0
\end{pmatrix},
\end{align}
or for open boundary conditions with
\begin{align}
A_{2n}^{s=-1}=\begin{pmatrix}
1 \\
0\end{pmatrix}&, 
A_{2n}^{s=1}=\begin{pmatrix}
0 \\
1\end{pmatrix}, \\
A_{2n-1}^{s=-1}=\begin{pmatrix}
1 & 0\end{pmatrix}&, 
A_{2n-1}^{s=1}=\begin{pmatrix}
0 & 1\end{pmatrix}.
\end{align}
Since this state is a MPS, it is also a SBS with 1 string. The RBM representation of the same state requires $N/2$ strings. In practice the number of non-zero coefficients are comparable, since in both cases the representation is sparse, but for numerical purposes a fully-connected RBM needs of the order $O(N^2)$ parameters before finding the exact ground state, while a MPS or SBS with one string will need $O(N)$ parameters for both open and periodic boundary conditions.

\begin{figure}[ht]
\centering
\includegraphics[width = 5cm]{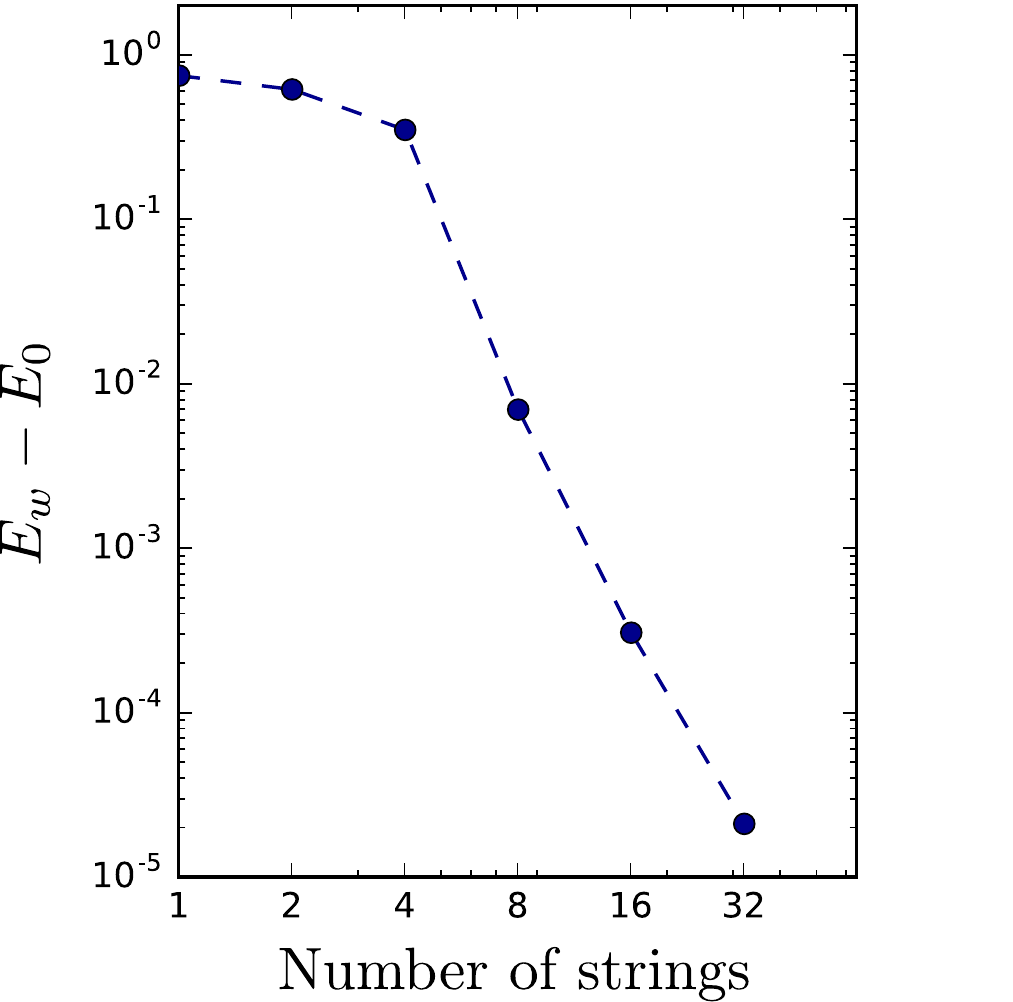}
\caption{Energy difference with the exact ground state energy of a spin-1 extension of a RBM (Eq.~\eqref{spin1rbm}) with $D=2$ and different number of strings for the AKLT model on a spin-1 chain with $8$ spins. A non-local SBS with non-commuting matrices and one string is exact within numerical accuracy.}
\label{fig:aklt}
\end{figure}

Another example is the AKLT model\cite{AKLT1987} defined by the following spin-$1$ Hamiltonian in periodic boundary conditions:
\begin{align}
H=\sum_{i=1}^{N} \left[\frac{1}{2} \mathbf{S}_i\cdot \mathbf{S}_{i+1}+\frac{1}{6} \left(\mathbf{S}_i\cdot \mathbf{S}_{i+1}\right)^2 + \frac{1}{3}\right].
\end{align} 
Its ground state has a simple form as a MPS of bond dimension $2$. It can also be written as an exact RBM by mapping the system to a spin-$1/2$ chain, but the number of hidden units needed for an exact representation scales as $O(N^2)$ in the system size\cite{XunGao}. We have numerically optimized the spin-1 extension of a RBM with form Eq.~\eqref{spin1rbm} (see  Appendix~\ref{optimization} for the details of the numerical optimization) and found that already for small sizes of the chain a much higher number of parameters is required to approach the ground state energy as compared to a SBS with non-commuting matrices, which is exact with one string of bond dimension 2 (Fig.~\ref{fig:aklt}). We will also show in Section \ref{section3} that in some other cases the RBM needs less parameters than a SBS to obtain a similar energy. This demonstrates that both RBM and SBS have advantages and that their efficiency depends on the particular model that is investigated. It remains an open question whether there exist MPS or SBS which can provably not be efficiently approximated by a RBM (for which the RBM would need exponentially many parameters).

To be able to use both the advantages of RBM (efficient to compute, few parameters) and of SBS (complex representation, geometric interpretation), one can use the flexibility of SBS by including some strings that have a full MPS over the whole lattice, some strings which include only local connections and that will ensure that the locality of the system is preserved, and some strings that have the form of an RBM and that can easily capture large entanglement and long-range correlations.
In many cases of interest, an initial approximation of the ground state can be obtained, either by optimizing simpler wave functions or by first applying DMRG to optimize a MPS. This initial approximation can then be used in conjunction with the previous Ansatz classes by multiplying an Ansatz wave function with the initial approximation. For the resulting wave function
\begin{align}
\psi_w(\mathbf{s})=\psi_w^{\text{init}}(\mathbf{s}) \psi_w^{\text{SBS}}(\mathbf{s}),\label{productwf}
\end{align}
the ratio of the wave function on two configurations as well as the log-derivatives depend only on the respective ratio and log-derivatives of each separate wave function, making the application of the VMC method straightforward. This procedure has the advantage of reducing the number of parameters necessary for obtaining a good approximation to the ground state and making the optimization procedure more stable, since the initial state is not a completely random state. Such a procedure provides a generic way to enhance the power of more specific Ansatz wave functions tailored to particular problems, as we will demonstrate in the next section. A similar technique has been used to construct tensor-product projected states with tensor networks in Ref.~\onlinecite{Sikora2015} and more generally it can be used to project the wave function of an initial reference state in a Fock space and is thus also suitable to describe fermionic systems.

\section{Application to chiral topological states}
\label{section3}

In this section we turn to a practical application on a challenging problem for traditional tensor-network methods, namely the approximation of a state with chiral topological order. While chiral topological PEPS have been constructed, the resulting states are critical. Moreover the local parent Hamiltonian of a chiral fermionic Gaussian PEPS has to be gapless\cite{Dubail2015}. In the following we investigate if this obstruction carries on to the tensor-network and neural-network states that we have introduced previously.

\subsection{RBM can describe a Laughlin state exactly}

Let us consider a lattice version of the Laughlin wave function at filling factor $1/2$ defined for a spin-$1/2$ system as
\begin{align}
\psi_{\text{Laughlin}}(\mathbf{s})=\delta_\mathbf{s} \prod_k \chi_k^{s_k}\chi\prod_{i<j} (z_i-z_j)^{\frac{1}{2} s_i s_j},\label{laughlinwf}
\end{align}
where $\delta_\mathbf{s}$ fixes the total spin to $0$, the $z_i$ are the complex coordinates of the positions of the lattice sites and the phase factor are defined as $\chi_k^{s_k}=e^{i\pi(k-1)(s_k+1)/2}$, ensuring that the state is a singlet. This wave function is equivalent to the Kalmeyer-Laughlin wave function in the thermodynamic limit and has been shown to describe a lattice state sharing the topological properties of the continuum Laughlin states on several lattices\cite{Nielsen2012,Tu2014,Glasser2016}. In addition, it can be written as a correlator from conformal fields, which has enabled the exact derivation of parent Hamiltonians for this state on any finite lattice\cite{Nielsen2011}.

The Laughlin wave function has the structure of a Jastrow wave function and we have shown in Section~\ref{jastrowrbmcorrespondence} that any Jastrow wave function can be written as a RBM with $M=N(N-1)/2$ hidden units. It follows that RBM and non-local SBS can represent a gapped chiral topological state exactly. This is in sharp contrast to local tensor-network states for which there is no exact description of a (non-critical) chiral topological state known. This difference is due to the non-local connections in the RBM and Jastrow wave function which allow them to easily describe a Laughlin state. We note that a chiral p-wave superconductor is another example of a gapped chiral topological state which has been recently written as a (fermionic) quasi-local Boltzmann Machine\cite{Huang2017}.

The previous construction is however not satisfactory in the sense that the RBM requires a number of hidden units scaling as $O(N^2)$, which is too high for numerical purposes on lattices which are not extremely small. We thus turn to the approximate representation of the Laughlin wave function using a RBM.

\subsection{Numerical approximation of a Laughlin state}

The lattice Laughlin wave function we consider has an exact parent Hamiltonian on a finite lattice\cite{Nielsen2011} defined as
\begin{align}
H_{\text{parent}}=\frac{2}{3}&\sum_{i\neq j} |w_{ij}|^2 \boldsymbol{S}_i\cdot \boldsymbol{S}_j + \frac{2}{3}\sum_{i\neq j \neq k} \bar{w}_{ij} w_{ik} \boldsymbol{S}_j\cdot \boldsymbol{S}_k \nonumber\\
&- \frac{2i}{3}\sum_{i\neq j \neq k}\bar{w}_{ij} w_{ik} \boldsymbol{S}_i \cdot (\boldsymbol{S}_j\times \boldsymbol{S}_k),
\end{align}
where $w_{ij}=\frac{z_i+z_j}{z_i-z_j}$ and $\boldsymbol{S}_j=(S_j^x,S_j^y,S_j^z)$ is the spin operator at site $j$. We specialize to the square lattice with open boundary conditions and minimize the energy of different wave functions with respect to this Hamiltonian by applying the VMC method presented in Section~\ref{vmc} with a Stochastic Reconfiguration optimization which is equivalent to the natural gradient descent\cite{Sorella2001,Neuscamman2012,Amari1998} (details of the numerical optimization can be found in Appendix~\ref{optimization}). Results are presented in Table~\ref{table:1}. 

We find that EPS with plaquettes of size up to $3\times 3$ have an energy difference with the Laughlin state of the order $10^{-2}$, which is better than a short-range RBM (denoted sRBM) on $3\times 3$ plaquettes and up to $M^\prime=4$ hidden units per plaquette, while the energy of a fully connected RBM with $M=2N$ hidden units is within $10^{-5}$ of the energy of the ground state. The resulting RBM uses much less hidden units than would be required for it to be exact, yet reaches an overlap of $99.99\%$ with the Laughlin wave function. This result shows that the fully-connected structure of the RBM is an advantage to describe this state and that EPS can be used instead of short-range RBM. We have moreover found that EPS are easier to optimize numerically than a short-range RBM: they are more stable, since each coefficient is considered separately, no exponentials or products that lead to unstable behavior are present and the derivatives have a very simple form (Eq.~\eqref{logdeps}).

\begin{table}[h]
\centering
\begin{tabular}{c c c} 
 \hline  \hline
 Ansatz & $(E_w-E_0)/N$ & $|\langle \psi_w | \psi_{\text{Laughlin}}\rangle|$ \\
 \hline
EPS $2 \times 2$ & $\ 4.3\times 10^{-2}\ $ & $46.10\%$ \\
EPS $3 \times 3$ & $\ 2.2\times 10^{-2}\ $ & $75.79\%$ \\
sRBM $M^\prime=\ 1$ & $\ 8.3\times 10^{-2}\ $ & $0.01\%$  \\
sRBM $M^\prime=\ 2$ & $\ 3.1\times 10^{-2}\ $ & $46.32\%$  \\ 
sRBM $M^\prime=\ 4$ & $\ 2.5\times 10^{-2}\ $ & $59.07\%$  \\ 
RBM $M=\ N$ & $\ 5.8\times 10^{-4}\ $ & $99.7\%$  \\
RBM $M=2N$ & $\ 1.1\times 10^{-5}\ $ & $99.99\%$  \\ 
 \hline  \hline
\end{tabular}
\caption{Energy per site difference with the ground state energy and overlap with the Laughlin state of different Ansatz wave functions optimized with respect to the Hamiltonian $H_{\text{parent}}$ on a $6 \times 6$ square lattice with open boundary conditions. sRBM have $M^\prime$ hidden units connected to all spins in each plaquette of size $3\times 3$, while RBM have $M$ hidden units connected to all spins of the lattice.}
\label{table:1}
\end{table}

\subsection{Numerical approximation of a chiral spin liquid}

The previous results indicate that RBM might be useful for approximating chiral topological states numerically, but are limited to relatively small sizes due to the non-local nature of the parent Hamiltonian, which includes interactions between all triplets of spins on the lattice. In Ref.~\onlinecite{Nielsen2013} a local Hamiltonian stabilizing a state in the same class as the Laughlin state was obtained by restricting $H_{\text{parent}}$ to local terms and setting the long-range interactions to zero. This leads to the Hamiltonian
\begin{align}
H_l=J \sum_{<i,j>} \boldsymbol{S}_i\cdot \boldsymbol{S}_j + J_\chi \sum_{<i,j,k>_\circlearrowleft}\boldsymbol{S}_i \cdot (\boldsymbol{S}_j\times \boldsymbol{S}_k), 
\end{align}
where $<i,j>$ indicates indices of nearest neighbours on the lattice and $<i,j,k>_\circlearrowleft$ indicates indices of all triangles of neighboring spins, with vertices labelled in the counter clockwise direction. We focus on the case $J=1, J_\chi=1$ for which the ground state of $H_l$ has above $98\%$ overlap with the Laughlin wave function (Eq.\eqref{laughlinwf}) on a $4\times 4$ lattice. We minimize the energy of different classes of states on a $4\times 4$ and $10\times 10$ square lattice with open boundary conditions. For optimizing wave functions with tens of thousands of parameters we use a batch version of Stochastic Reconfiguration which optimizes a random subset of the parameters at each iteration (see Appendix~\ref{optimization}). We consider several Ansatz wave functions including EPS with plaquettes of size $2\times 2$, $3\times 2$, $4\times 2$ and $3 \times 3$, local SBS covering the lattice with horizontal, vertical and diagonal strings and increasing bond dimension, RBM with increasing number of hidden units, non-local SBS with diagonal matrices (denoted dSBS) or with non-commuting matrices of bond dimension $2$ and different number of strings covering the full lattice. We observe that while the optimization of EPS and SBS is particularly stable, the optimization of RBM can lead to numerical instabilities that are resolved by writing the RBM in the form presented in Eq.\eqref{RBMlog}. Since we use the same optimization procedure for all Ansatz wave functions and since the required time (and memory) to perform the optimization is mainly a function of the number of parameters and of the accuracy, we can compare the Ansatz classes by comparing how many parameters are needed to obtain a similar energy.

\begin{figure}[ht]
\centering
\subfloat[4x4 lattice]{\includegraphics[width = 6.8cm]{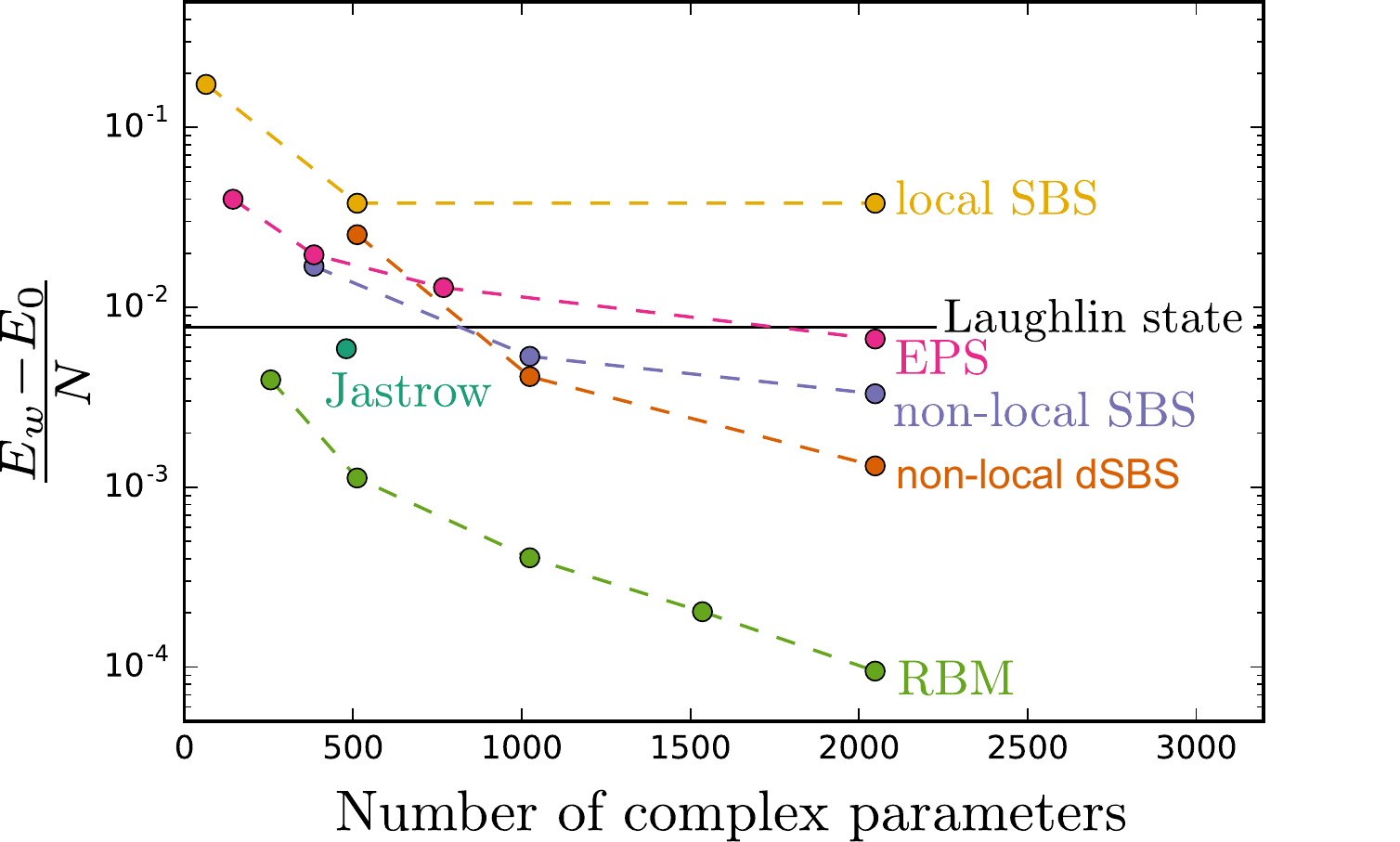}\label{results4x4i}}\\
\subfloat[10x10 lattice]{\includegraphics[width = 6.8cm]{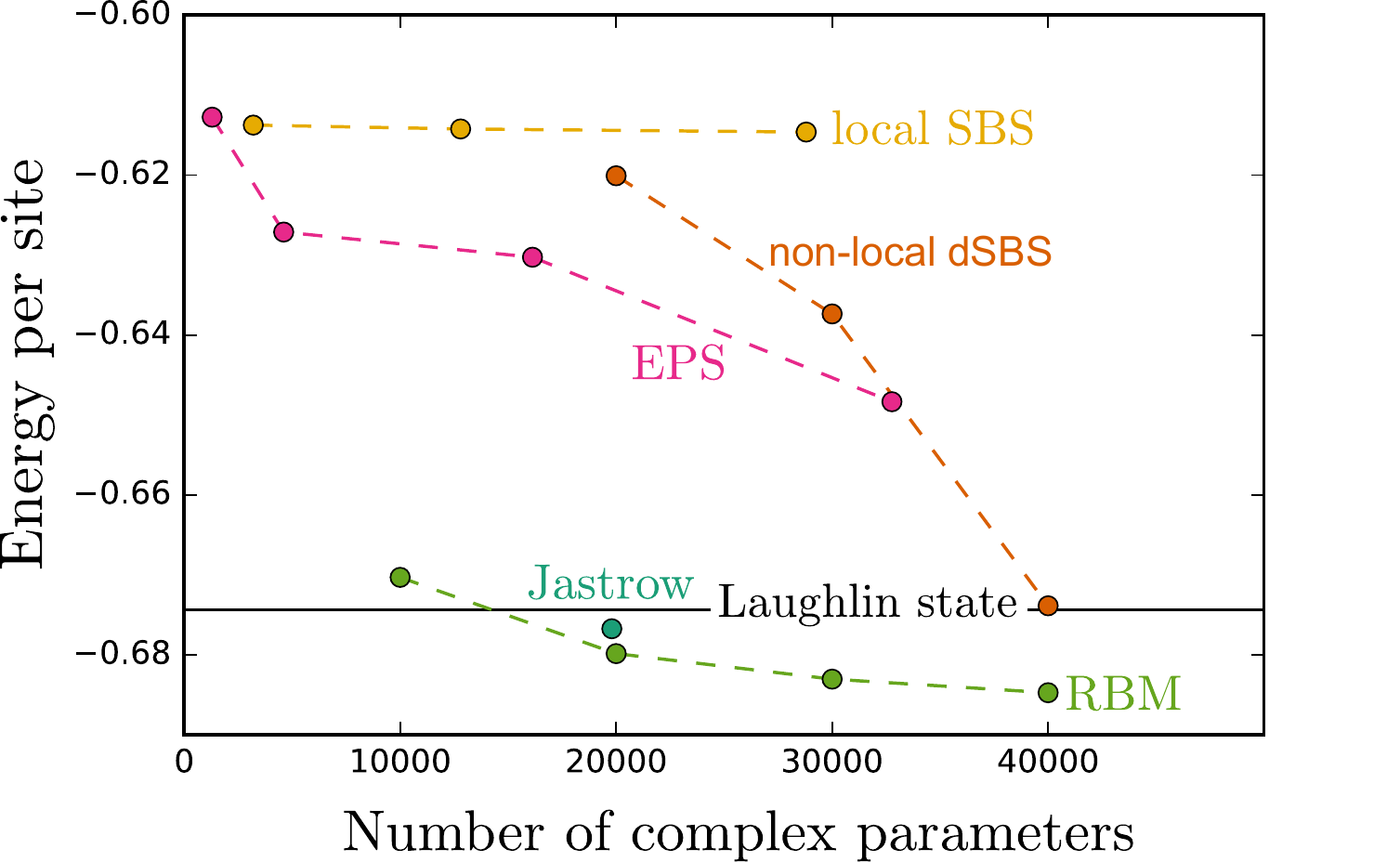}\label{results10x10i}}\\
\subfloat[10x10 lattice]{\includegraphics[width = 6.8cm]{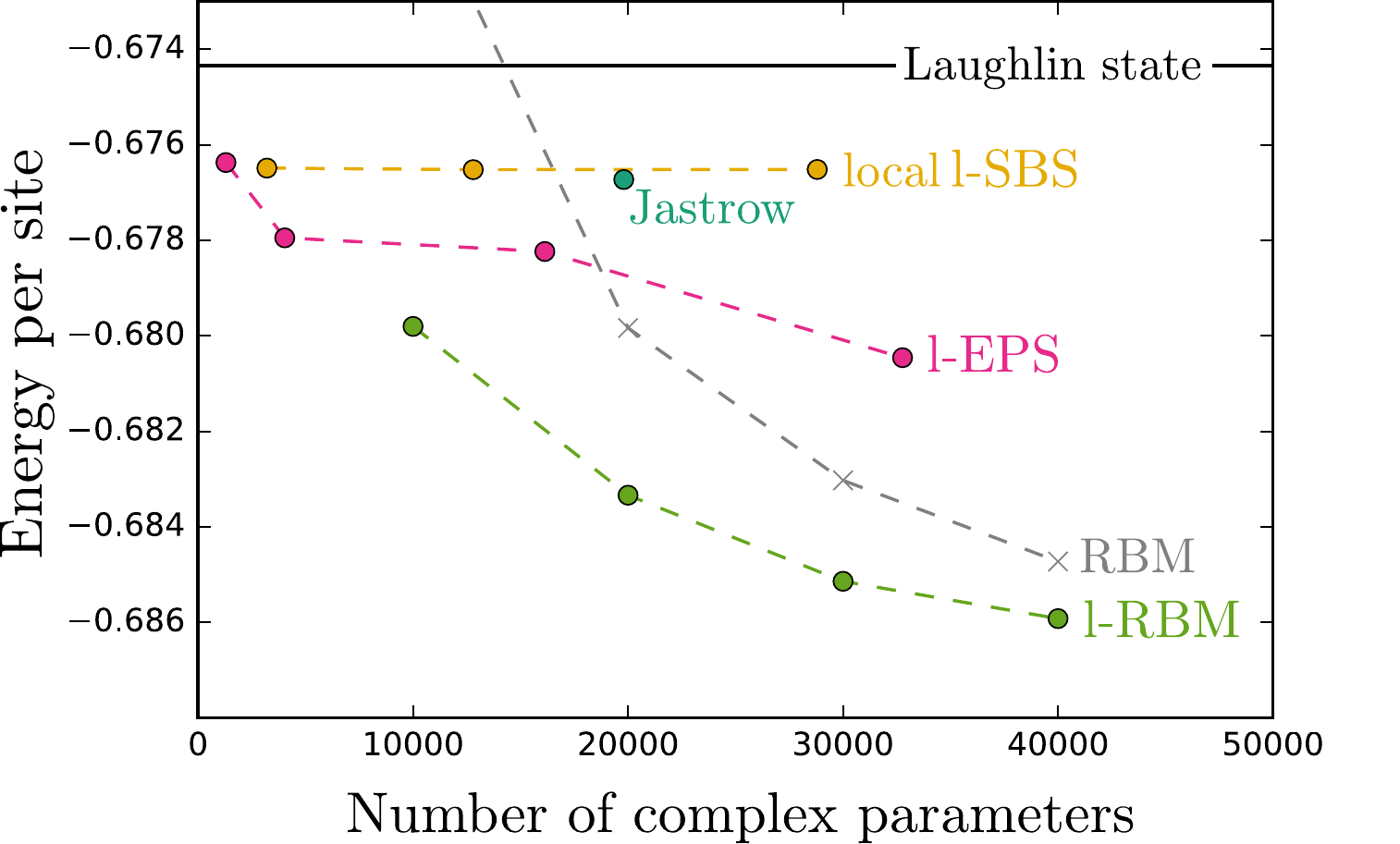}\label{results10x10ilaughlin}}
\caption{Energy of $H_l$ per site for different optimized Ansatz wave functions on a square lattice. The number of parameters ($N_p$) is modified by increasing the bond dimension D (local SBS, $N_p\propto D^2$), the size of the plaquettes (EPS, $N_p\propto M_P 2^P$, where $M_P$ is the number of plaquettes and $P$ is the number of spins in one plaquette), the number of strings $M_S$ (non-local SBS and dSBS, $N_p\propto M_S$) or the number of hidden units $M_h$ (RBM, $N_p\propto M_h$). (a) 4x4 lattice for which the energy difference with the exact ground state energy is plotted. (b) 10x10 lattice for which the exact ground state energy is unknown and the reference energy of the Laughlin state is indicated as a black line. (c) Optimization of wave functions that have been multiplied by the Laughlin wave function on a 10x10 lattice. The original RBM results are indicated for reference as grey crosses.}
\end{figure}

We first focus (Fig.~\ref{results4x4i}) on a $4\times 4$ lattice for which the exact ground state can be obtained using exact diagonalization. Local SBS have an energy higher than the Laughlin state and the energy is saturated with increasing bond dimension, which means that the pattern of horizontal, vertical and diagonal strings is not enough to capture all correlations in the ground state. While a large $4\times 4$ plaquette would make EPS exact on this small lattice, this would require $2^{16}$ parameters. The energy of the Laughlin state is already reached for $3 \times 3$ plaquettes. RBM with a number of hidden units larger than $N$ and non-local SBS with a corresponding number of strings have lower energy than the Laughlin state or the Jastrow wave function. When the number of strings grows, the energy decreases even further. On a larger $10 \times 10$ lattice (Fig.~\ref{results10x10i}) the exact ground state energy is unknown but we can compare the energy of the different Ansatz wave functions and observe similar results. Only the Jastrow wave function, non-local SBS and RBM have an energy comparable to the Laughlin state. Notice that non-local SBS have a constant factor more parameters than a RBM for the same number of strings. On the one side this allows SBS to achieve better energy than RBM with the same number of strings. On the other side this comes with the drawback than we can only optimize fewer strings and on the large lattice we are numerically limited to non-local dSBS with up to N strings. We can conclude that RBM are particularly efficient in this example since they require significantly less parameters than SBS for attaining the same energy. This has to be contrasted with the previous examples of the Majumdar-Gosh and AKLT models where the opposite was true. Therefore each class of states has advantages and drawbacks depending on the model we are looking at. We note in addition that a non-local SBS can be initialized with the results of a previous optimization with a RBM, which could provide a way of minimizing the difficulties of optimizing large number of parameters.

\begin{figure}[t]
\centering
\includegraphics[width = 4.8cm]{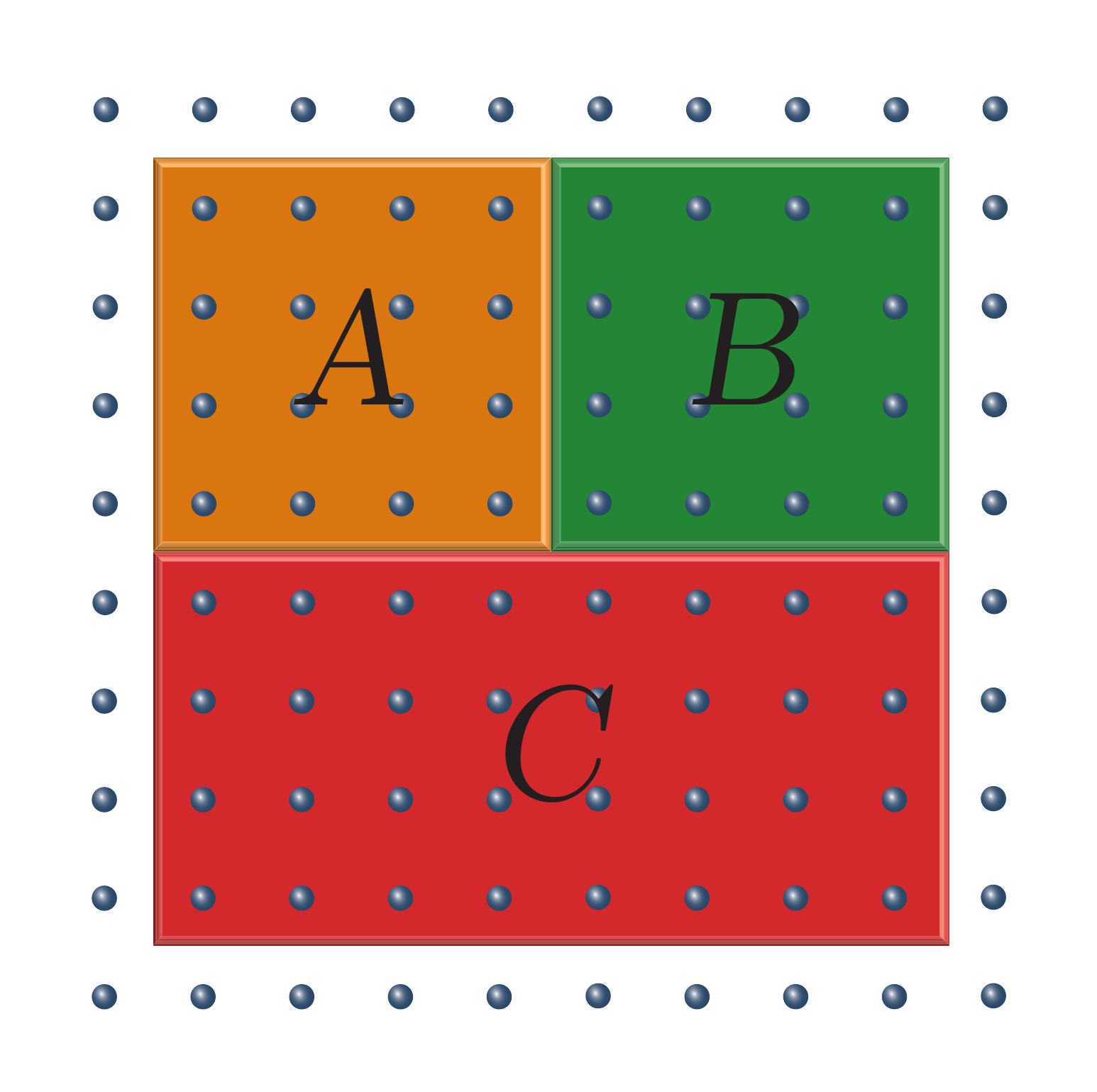}
\caption{Partition of the lattice used to compute the topological entanglement entropy.}
\label{TEEpartition}
\end{figure}

\begin{table}[ht]
\centering
\begin{tabular}{c c c} 
 \hline  \hline
 Ansatz & TEE \\
 \hline
Laughlin & $-0.339(3)$  \\ 
l-EPS $3 \times 3$& $-0.36(1)$ \\
RBM $M=4N$ & $-0.34(1)$ \\
l-RBM $M=4N$ & $-0.34(1)$  \\

 \hline  \hline
\end{tabular}
\caption{Topological entanglement entropy (TEE) of the analytical Laughlin state and optimized l-EPS, RBM and l-RBM.}
\label{table:2}
\end{table}

\begin{figure}[ht]
\centering
\subfloat[]{\includegraphics[width = 4.2cm]{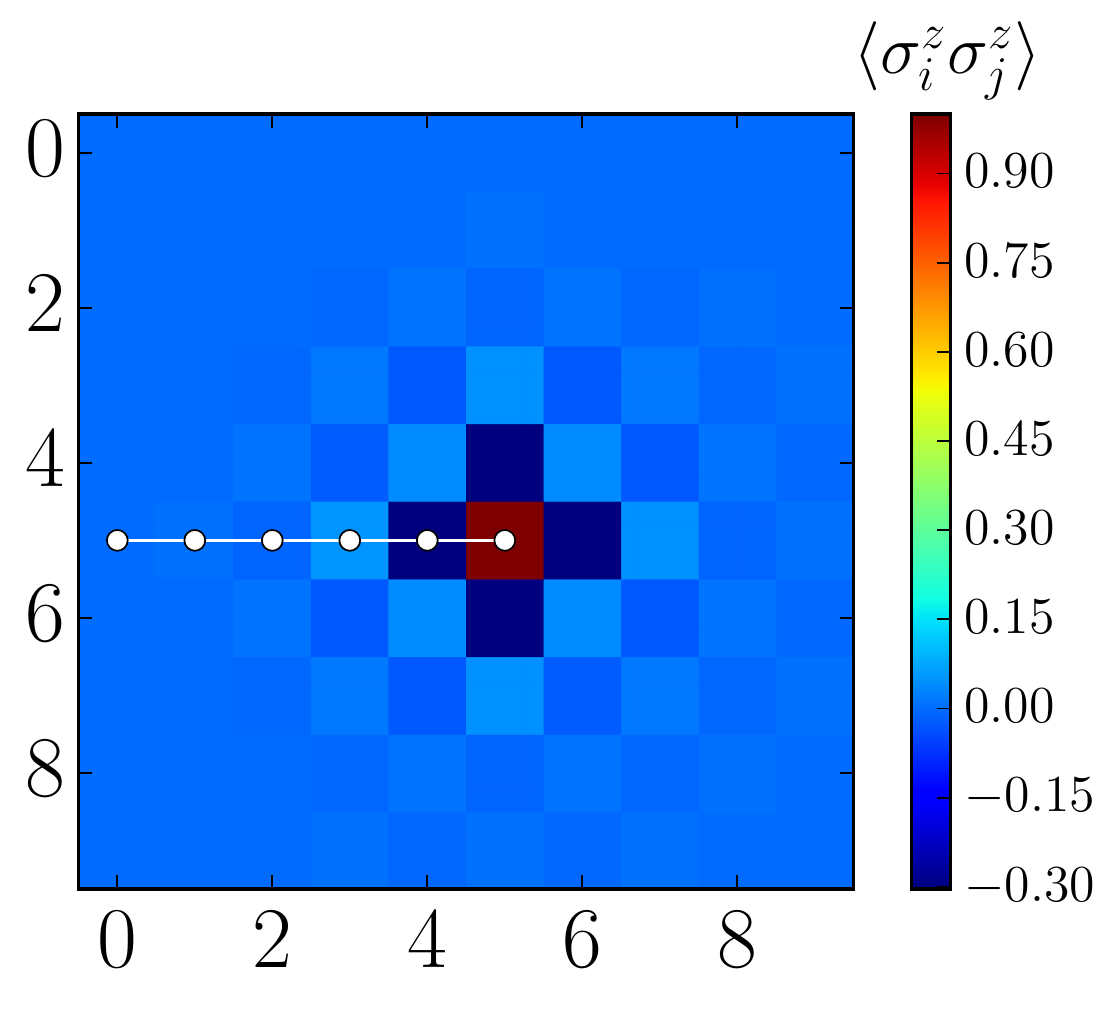}\label{correlations}}
\subfloat[]{\includegraphics[width = 3.5cm]{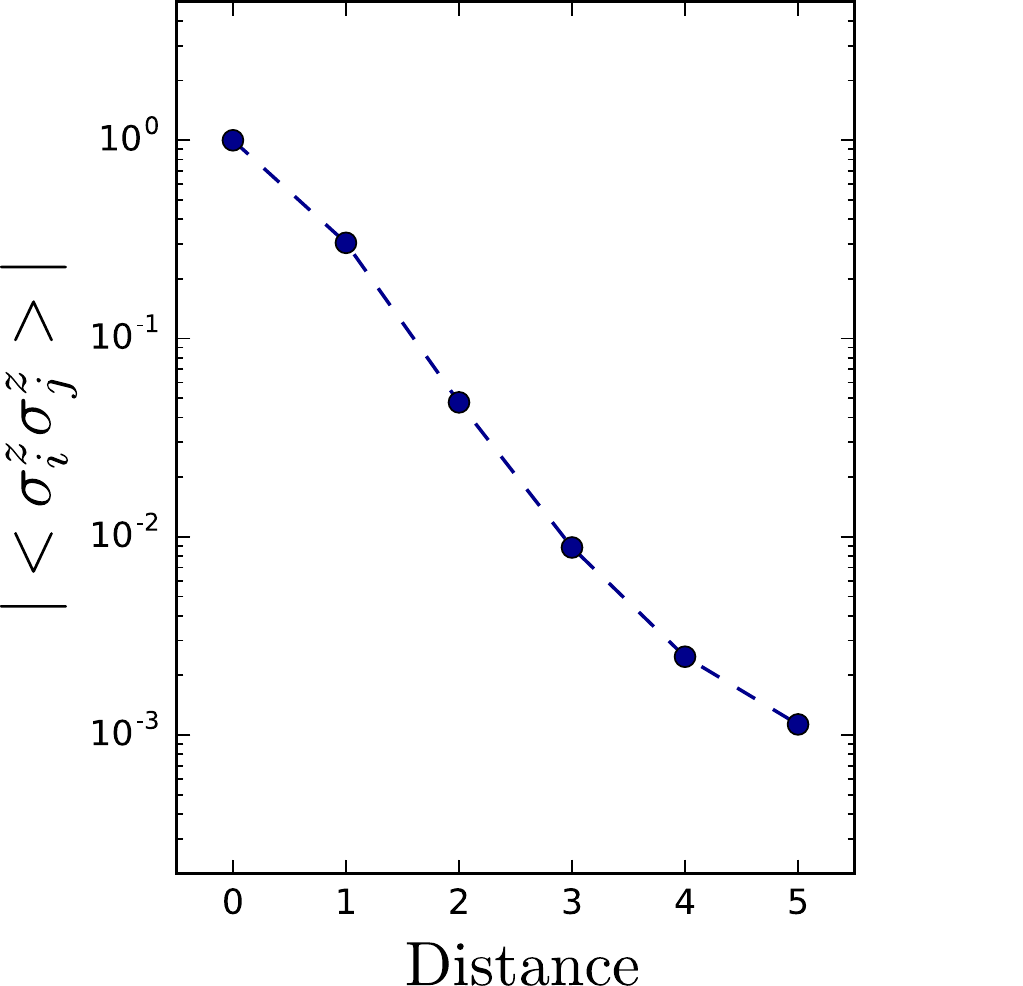}\label{correlations2}}
\caption{(a) The spin-spin correlation function between one lattice site (in red) and all other spins on the lattice measured on the optimized l-RBM with lowest energy reveals the antiferromagnetic behavior of the correlations. (b) Decay of the correlations with the distance across the direction indicated in (a) as a white solid line. The error bars are within dot size and finite size effects can already be seen for the last point.}
\end{figure}

As we have previously noticed, we can also use an initial approximation of the ground state in combination with the previous Ansatz classes. In the case of the Hamiltonian $H_l$, the analytical Laughlin wave function can be used as our initial approximation in Eq.~\ref{productwf}. We denote l-EPS (resp. l-SBS, l-RBM) a wave function that consists in a product of the Laughlin wave function and an EPS (resp. SBS, RBM) and minimize the energy of the resulting states. This allows us to obtain lower energies for each Ansatz class (Fig.~\ref{results10x10ilaughlin}). Once the wave functions are optimized, their properties can be computed using Monte Carlo sampling. To check that the ground state is indeed in the same class as the Laughlin state, we compute the topological entropy of some of the optimized states by dividing the lattice into four regions (Fig.~\ref{TEEpartition}) and computing the Renyi entropy $S^{(2)}_A = - \ln {\rm Tr} \;  \rho^2_A$ of each subregion using the Metropolis-Hastings Monte Carlo algorithm with two independent spin chains \cite{Hastings2010,Wildeboer2015}. The topological entanglement entropy is then defined as\cite{Kitaev2006,Levin2006}
\begin{align}
S_{\text{topo}}=&S^{(2)}_A+S^{(2)}_B+S^{(2)}_C-S^{(2)}_{AB}\nonumber\\
&-S^{(2)}_{AC}-S^{(2)}_{BC}+S^{(2)}_{ABC},
\end{align}
and is expected to be equal to $-\ln 2\approx -0.347$ for the Laughlin state\cite{Zozulya2007}. The results we obtain are presented in Table~\ref{table:2} and provide additional evidence that the ground state of $H_l$ has the same topological properties as the Laughlin state. The Hamiltonian $H_l$ was recently investigated on an infinite lattice using infinite-PEPS\cite{Poilblanc2017} and further evidence was provided that the ground state is chiral. The PEPS results suggest the presence of long-range algebraically decaying correlations that may be a feature of the model or a restriction of PEPS to study chiral systems. The correlations on short distances agree with the correlations that we can compute on our finite system (Fig.~\ref{correlations}) but our method does not allow us to make claims about the long-distance behavior of the correlation function. We also observe that fully-connected RBM cannot be defined directly in the thermodynamic limit without a truncation of the distance of the interaction between visible and hidden units, thus transforming the RBM into a short-range RBM (albeit of larger range than an EPS). In Ref.~\cite{Deng2017} it was observed that the entanglement entropy of some specific short-range RBM can be computed analytically from the weights of the RBM. The method we use here works in the general case and also for a fully-connected RBM, but requires Monte Carlo sampling of the wave function. The optimized RBM weights encode every information about the wave function, it would thus be interesting to understand more precisely which quantities can be extracted directly from them. Whether direct information about the phase of the system can be obtained in this way without requiring Monte Carlo sampling remains an interesting open problem for future work.

\bigskip
\section{Conclusion}
\label{SEC:Conclusion}
\bigskip
We have shown that there is a strong connection between Neural-Network Quantum States in the form of Boltzmann Machines and some Tensor-Network states that can be optimized using the Variational Monte Carlo method : while short-range Restricted Boltzmann Machines are a subclass of Entangled Plaquette States, fully connected Restricted Boltzmann Machines are a subclass of String-Bond States. These String-Bond States are however different from traditional String-Bond States due to their non-local structure which connects every spin on the lattice to every string. This enabled us to generalize Restricted Boltzmann Machines by introducing non-local (diagonal or non-commuting) String-Bond States which can be defined for larger local Hilbert space and with additional geometric flexibility. We compared the power of these different classes of states and showed that while there are cases where String-Bond States require less parameters than fully-connected Restricted Boltzmann Machines to describe the ground state of a many-body Hamiltonian, there are also cases where the additional parameters in each string make String-Bond States less efficient to optimize numerically. We applied these methods to the challenging problem of describing states with chiral topological order, which is hard for traditional Tensor Networks. We showed that every Jastrow wave function, and thus a Laughlin wave function, can be written as an exact Restricted Boltzmann Machine. In addition we gave numerical evidence that a Restricted Boltzmann Machine with a much smaller number of hidden units can still give a good approximation to the Laughlin state. Finally we turned to the approximation of the ground state of a chiral spin liquid and showed that Restricted Boltzmann Machines achieve a lower energy than the Laughlin state and the same topological entanglement entropy. We argued that combining different classes of states allows to take advantage of the initial knowledge of the model and of the particularities of each class. This was demonstrated by combining a Jastrow wave function to Tensor Networks and Restricted Boltzmann Machines, which allowed us to get lower energies than the initial states and characterize the ground state. 

Our work sheds some light on the representative power of Restricted Boltzmann Machines and establish a bridge between their optimization and the optimization of Tensor Network states. On the one hand, the methods developed in this work can be used to target the ground state of other Hamiltonians and it would be interesting to know whether similar results can be achieved for example for non-Abelian chiral spin liquids\cite{Greiter2009,Glasser2015} or generalized to fermionic systems of electrons in the continuum displaying the Fractional Quantum Hall effect. On the other hand, we also showed that some tools used in machine learning can be rephrased in Tensor Network language, thus providing additional physical insights about the systems they describe. Matrix Product States have already been used as a tool for supervised learning\cite{Novikov2016,Stoudenmire2016} and our work opens up the possibility of using not only Restricted Boltzmann Machines, but also String-Bond States to represent a probability distribution over some data while encoding additional information about its geometric structure.

\textit{Note added.} After the completion of this manuscript, related independent work came to our attention. Y. Nomura et al.\cite{Nomura2017} combine RBM with pair product wave functions and apply them to the Heisenberg and Hubbard models. S. R. Clark\cite{Clark2017} constructs a mapping between RBM and EPS/Correlator Product States. R. Kaubruegger et al.\cite{Kaubruegger2017} give further analytical and numerical evidence supporting the application of RBM to chiral topological states such as the Laughlin state.

\begin{acknowledgments}
We would like to thank Martin Ganahl, Xun Gao and Giuseppe Carleo for discussions. This work was supported by the ERC grant QUENOCOBA, ERC-2016-ADG (Grant no. 742102). NP and MA acknowledge financial support from ExQM.
\end{acknowledgments}
\bigskip

\newpage
\allowdisplaybreaks
\appendix

\section{Jastrow wave functions are Restricted Boltzmann Machines}
\label{Appjastrowasrbm}
Let us show that a RBM with one hidden unit can represent any function $f$ of two spins. It then follows that a RBM with $M=N(N-1)/2$ hidden units, each representing a function of one pair of spins, can represent a Jastrow wave function. We parametrize $f$ by its four values on two spins $s_1,s_2\in\{-1,1\}$ and solve for a system of four non-linear equations:
\begin{align}
F_{11}&=A_1 A_2 \left(W_1 W_2+\frac{1}{W_1 W_2}\right)\\
F_{-1-1}&=\frac{1}{A_1 A_2} \left(W_1 W_2+\frac{1}{W_1 W_2}\right)\\
F_{1-1}&=\frac{A_1}{ A_2} \left(\frac{W_1}{ W_2}+\frac{W_2}{ W_1}\right)\\
F_{-11}&=\frac{A_2}{ A_1} \left(\frac{W_2}{ W_1}+\frac{W_1}{ W_2}\right),
\end{align}
where we have set $B_1=B_2=1$. The RBM is well defined when all parameters are non-zero and we change of variables by defining $X=W_1 W_2$, $Y=\frac{W_1}{ W_2}$, $A=A_1 A_2$, $B=\frac{A_1}{ A_2}$, obtaining a new set of equations:
\begin{align}
F_{-1-1} A^2 &= F_{11}\\
F_{-11} B^2 &= F_{1-1}\\
X^2-\frac{1}{A}X+1&=0\\
Y^2-\frac{1}{B}Y+1&=0.
\end{align}
We first suppose that the values $F_{s_i s_j}$ are non-zero. These quadratic equations all have non-zero analytical solutions in the complex plane, that we denote $A_0$, $B_0$, $X_0$, $Y_0$. The original parameters are then the solutions of
\begin{align}
W_1^2 &= X_0 Y_0\\
W_2^2 &= X_0 / Y_0 \\
A_1^2&= A_0 B_0\\
A_2^2&=A_0 / B_0,
\end{align}
which is again a set of quadratic equations with non-zero analytical solutions. If $F_{11}=F_{-1-1}=0$ (resp. $F_{1-1}=F_{-11}=0$), the exact solution is given directly by $A_0=1, X_0=i$ (resp. $B_0=1, Y=i$). In the remaining cases where some $F_{s_i s_j}$ are zeros, the equations do not always have an exact solution, but the function can still be approximated to arbitrary precision. This case corresponds to strong restrictions on the part of the Hilbert space which is used to write the wave function and these constraints can also be imposed on the states directly by adding a delta function to the wave function which is equal to $1$ only when the constraints on the spins are satisfied. Having a Markov Chain Monte Carlo sampling which does not visit these states then allows for a more efficient sampling.

\section{Optimization procedure}
\label{optimization}
The goal is to minimize the energy $E_{\mathbf{w}}$ depending on some vector of parameters $\mathbf{w}$. We define $\mathbf{f}$ to be the energy gradient vector at $\mathbf{w}$. Expanding the energy to first order around $\mathbf{w}$ leads to the steepest gradient descent, where the variational parameters are updated at each iteration according to $\mathbf{w'}=\mathbf{w}+\boldsymbol{\gamma}$, with a change of parameters given by $\boldsymbol{\gamma}=-\alpha \mathbf{f}$. Here $\alpha$ is a small step size. Expanding the energy to second order instead would result in the Newton method with a change of parameters given by: 
\begin{align}
\boldsymbol{\gamma}=-\alpha \mathbf{H}^{-1}\mathbf{f},\label{newtonmethod}
\end{align}
where $\mathbf{H}$ is the Hessian of the energy. Small changes of the variational parameters may however lead to big changes in the wave function, especially in the case of compact non-local representations like RBM in which each parameter affects each part of the wave function. Taking into account the metric of changes of the wave function leads to the Stochastic Reconfiguration\cite{Sorella2001} method, which is equivalent to the natural gradient descent\cite{Amari1998} and replaces the Hessian in Eq.~\eqref{newtonmethod} by the covariance matrix of the derivatives of the wave function, avoiding the computation of the second-order derivatives of the energy. 

The Stochastic Reconfiguration method can also be viewed as an approximate imaginary-time evolution in the tangent space of the wave function. Consider the normalized wave function $|\bar \psi_0\rangle$ and its derivatives
\begin{align}
|\bar \psi_0\rangle&=\frac{|\psi_0\rangle}{\sqrt{\langle \psi_0|\psi_0\rangle}},\\
|\bar \psi_i\rangle&=\frac{|\psi_i\rangle}{\sqrt{\langle \psi_0|\psi_0\rangle}}-\frac{\langle \psi_0|\psi_i\rangle}{\langle \psi_0|\psi_0\rangle}\frac{|\psi_0\rangle}{\sqrt{\langle \psi_0|\psi_0\rangle}},
\end{align}
defining a non-orthogonal basis set $\Omega$. Expanding the wave function to linear order around some parameters $\mathbf{w}$ leads to 
\begin{align}
|\bar \psi(\mathbf{w}+\boldsymbol{\gamma})\rangle=\sum_{i=0}^{N_w} \gamma_i |\bar \psi_i\rangle.
\end{align}
To minimize the energy, one can apply the imaginary-time evolution operator $e^{-\alpha H}$, which expanded to first order for small $\alpha$ is $1-\alpha H$. The change of coefficients $\boldsymbol{\gamma}$ is found by applying this operator to $|\bar \psi(\mathbf{w}+\boldsymbol{\gamma})\rangle$ and projecting in the set $\Omega$, which leads to the equation
\begin{align}
-\alpha \langle \bar \psi_i |H| \bar \psi_0 \rangle =\sum_{j=1}^M \langle \bar \psi_i |\bar \psi_j\rangle \gamma_j,
\end{align}
which can be rewritten as
\begin{align}
\boldsymbol{\gamma}=-\alpha \mathbf{S}^{-1}\mathbf{f}\label{updateparameters},
\end{align}
where $S_{ij}=\langle \bar \psi_i |\bar \psi_j\rangle$ and $f_i=\langle\bar \psi_i |H| \bar \psi_0\rangle$.
If we expand these expressions as expectation values over the probability distribution $p(\mathbf{s})=\frac{|\psi_w(\mathbf{s})|^2}{\sum_\mathbf{s} |\psi_w(\mathbf{s})|^2}$, we obtain
\begin{align}
f_i&=\langle\Delta_i^* E_{\text{loc}}\rangle-\langle\Delta_i^*\rangle \langle E_{\text{loc}}\rangle,\\
S_{ij}&=\langle\Delta_i^* \Delta_j\rangle-\langle\Delta_i^*\rangle \langle \Delta_j\rangle,
\end{align}
where the local energy is defined as $E_{\text{loc}}(\mathbf{s})=\sum_{\mathbf{s}'}\langle \mathbf{s}|H|\mathbf{s}'\rangle \frac{\psi_w(\mathbf{s}')}{\psi_w(\mathbf{s})}$ and the log-derivative of the wave function as $\Delta_w(\mathbf{s})=\frac{1}{\psi_w(\mathbf{s})}\frac{\partial \psi_w(\mathbf{s})}{\partial w}$. Finally, the complete algorithm is as follows:
\begin{enumerate}
\item Using a Metropolis-Hastings algorithm, generate samples of the probability $p$ and compute stochastic estimates for the expectation values $\langle \Delta_j\rangle$, $\langle E_{\text{loc}}\rangle$, $\langle\Delta_i^* E_{\text{loc}}\rangle$, $\langle\Delta_i^* \Delta_j\rangle$
\item Construct the vector $\mathbf{f}$ and matrix $\mathbf{S}$,
\item Update the parameters according to $\mathbf{w}\leftarrow \mathbf{w}-\alpha \mathbf{S}^{-1}\mathbf{f}$,
\item Repeat the full procedure until convergence of the energy.
\end{enumerate}
In practice we repeat the full procedure $1000$ to $20000$ times until the energy is converged. To optimize a large number of parameters we randomly select a subset of the parameters of size up to $10000$ at each iteration of the algorithm and update only these parameters. This reduces the computational cost associated with the operations dealing with $\mathbf{f}$ and $\mathbf{S}$. Moreover we can avoid forming the full matrix $\mathbf{S}$ by instead solving Eq.~\eqref{updateparameters} with a conjugate-gradient solver \cite{Neuscamman2012}. Numerical stability can be achieved by adding a small constant $\epsilon$ to the diagonal elements of the matrix $\mathbf{S}$, rotating the direction of change towards the steeped descent direction. We find that a step size $\alpha$ of the order $1/\sqrt{i}$, where $i$ is the iteration step, works well in conjunction with a large stabilization at the beginning, while a fixed step size can also be chosen in conjunction with a small stabilization of the order $10^{-4}-10^{-8}$ by performing several optimizations. At the later stages of the optimization, the step size is lowered to ensure that the energy is converged. Further improvements are achieved by projecting the wave functions in a subset of fixed total spin when it is conserved by the Hamiltonian we consider\cite{Tahara2008}. The spin-flip symmetry can be enforced in a RBM by choosing the bias $b_i=0$.

\bibliographystyle{apsrev4-1-title}
\bibliography{bibliorbm}
\end{document}